\documentclass[aps,prc,
twocolumn,amsmath,amssymb,
]{revtex4-2}
\usepackage[english]{babel}
\usepackage{graphicx}
\usepackage{dcolumn}
\usepackage{bm}

\usepackage[dvipsnames]{xcolor}

\begin{document}



\title{Measurement of the $^{236}$U fission cross section and angular distributions of fragments from fission of $^{235}$U and $^{236}$U in the neutron energy range of 0.3--500 MeV}

\author{A.S. Vorobyev}
\email[]{vorobyev\underline{\phantom{0}}as@pnpi.nrcki.ru}
\affiliation{B.P. Konstantinov Petersburg Nuclear Physics Institute of National Research Center ``Kurchatov institute'', 188300 Gatchina, Leningradskaya oblast, Russia}
\author{A.M. Gagarski}
\affiliation{B.P. Konstantinov Petersburg Nuclear Physics Institute of National Research Center ``Kurchatov institute'', 188300 Gatchina, Leningradskaya oblast, Russia}
\author{O.A. Shcherbakov}
\affiliation{B.P. Konstantinov Petersburg Nuclear Physics Institute of National Research Center ``Kurchatov institute'', 188300 Gatchina, Leningradskaya oblast, Russia}
\author{L.A. Vaishnene}
\affiliation{B.P. Konstantinov Petersburg Nuclear Physics Institute of National Research Center ``Kurchatov institute'', 188300 Gatchina, Leningradskaya oblast, Russia}
\author{A.L. Barabanov}
\affiliation{National Research Center ``Kurchatov institute'', 123182 Moscow,  Russia,}
\affiliation{National Research Nuclear University MEPhI, 115409 Moscow, Russia,}
\affiliation{Moscow Institute of Physics and Technology, 141701 Dolgoprudny, Moscow Region, Russia}
\author{T.E. Kuz'mina}
\affiliation{JSC ``V.G. Khlopin Radium Institute'', Rosatom State Atomic Energy Corporation, 194021 St. Petersburg, Russia}

\date{\today}

\begin{abstract}
The $^{236}$U fission cross section and the angular distributions of fragments from fission of $^{235}$U and $^{236}$U were measured for incident neutron energies from 0.3 MeV to 500 MeV on the time-of-flight spectrometer of the neutron complex GNEIS at the NRC ``Kurchatov Institute'' -- PNPI. Fission fragments were registered using position-sensitive low-pressure multiwire counters. In the neutron energy range above 20 MeV, the angular distributions of $^{236}$U fission fragments were measured for the first time. The fission cross section of $^{236}$U$(n,f)$ was measured relative to the fission cross section of $^{235}$U$(n,f)$, which is an accepted international standard. The obtained data are compared with the results of other experimental works. Theoretical calculations of the fission cross section and the anisotropy of angular distribution of fission fragments for the $^{236}$U$(n,f)$ reaction performed within the framework of our approach are presented and discussed.
\end{abstract}



\maketitle


\section*{Introduction}

At present, more accurate data on nuclear fission are needed for the design of modern and future nuclear reactors \cite{Gen4_2014}, for the development of methods for controlling the non-proliferation of nuclear materials, non-destructive testing methods used at nuclear power plants and fuel reprocessing plants, as well as for the design of Accelerator-Driven Systems (ADS), including powerful neutron sources~\cite{ADS_2015}. The advantage of ADS installations lies not only in the safety of control compared to traditional nuclear reactors, but also in the possibility of using even-even isotopes of uranium, plutonium, thorium, which are not fissile by thermal neutrons as fuel in such systems, as well as the transmutation of highly toxic waste from traditional nuclear power. For the scientific and technological substantiation of this direction of nuclear energy development, it is necessary to create a reliable nuclear database, primarily neutron data, in the field of intermediate and high energies.

Obtaining such nuclear data is an urgent and at the same time extremely difficult task. At the same time, the accuracy of calculations performed using existing theoretical approaches does not meet the needs of new nuclear technologies. Therefore, in order to reduce the total uncertainty of the estimated data, it is necessary to measure the fission cross sections in a wider range of neutron energies using various experimental techniques, since this is the only way to detect possible systematic errors. It is also important to note that obtaining new nuclear data will not only fill in the existing gaps in the experimental database, but also stimulate the creation of theoretical models used both for analyzing experimental results and for engineering calculations.

The desire to reduce the cost of nuclear power plant designs with fast neutron reactors imposes increased requirements for the required burnout depth of fuel loaded into the reactor, which leads to an increase in the amount of generated fission products and transactinium nuclei ($^{240}$Pu, $^{241}$Pu, $^{242}$Pu and $^{241}$Am). As a result, the role of these nuclides in the reactor and in the external fuel cycle increases, and the accuracy of the nuclear data required for them becomes comparable to the accuracy required for the main fissile and fuel isotopes. Today, as an alternative to the uranium-plutonium fuel cycle, the possibility of using a thorium-uranium nuclear fuel cycle is being considered. In this case, in addition to the main isotopes $^{232}$Th, $^{233}$U and $^{234}$U, the nucleus $^{236}$U is of particular importance, which is produced in a fuel with high burnout, and from which, as a result of sequential radiation capture, $^{237}$Np and $^{238}$Pu nuclei are formed, and as a result of the $(n,2n)$ reaction --- the $^{235}$U nucleus.

The data available in the literature on the neutron-induced fission cross section of the $^{236}$U were obtained both using beams of quasi-monoenergetic neutrons formed at Van de Graaf accelerators~\cite{Meadows_1978, Fursov_1985, Goverdovskii_1985, Kanda_1986, Meadows_1988, Diakaki_2020} and neutron sources with a continuous spectrum using the time-of-flight method~\cite{Behrens_1977, Lisowski_1992, Sarmento_2011, Tovesson_2014, Ren_2020}. Most of these data relate to neutrons with energies below 40 MeV, and only in two works \cite{Lisowski_1992, Tovesson_2014} the incident neutron energy range was expanded to 400 MeV and 200 MeV, respectively.

In \cite{Diakaki_2020}, measurements were carried out using Micromegas detectors, whereas in the rest of these studies, the registration of fission fragments was carried out by multi-section ionization chambers. At the same time, the remark of authors in Ref.~\cite{Goverdovskii_1985} that at neutron energies of 4-10 MeV the spread of experimental data is $\approx 8\,\%$, which is higher than their stated accuracy, remains relevant. It’s ought to mention that the required accuracy is $\approx 5\,\%$ \cite{Pronyaev_1999}.

In all the experimental works mentioned above, the fission cross section of the $^{236}$U nucleus was measured relative to the fission cross section of the $^{235}$U nucleus, which is an international standard \cite{Marcinkevicius_2015, Carlson_2018}. This makes it possible to minimize errors associated with the uncertainty of the neutron flux. It should also be noted that the current methods \cite{Meadows_1978, Fursov_1985, Goverdovskii_1985, Meadows_1988, Behrens_1977} for measuring the threshold fission cross sections also allow minimizing the error caused by the uncertainty of the mass of the samples under study.

As for the data on the differential cross-section of the $^{236}$U neutron-induced fission, or, in other words, on the angular distribution of fragments relative to the reaction axis, there are very few of them. The energy dependence of the angular anisotropy of fission fragments in the $^{236}$U$(n,f)$ reaction was studied only in \cite{Simmons_1960, Leachman_1965, Huizenga_1969, Shpak_1991} for neutrons with energies from 0.4 to 15 MeV.

In this paper, we set ourselves the goal of measuring both the fission cross section and the angular distribution of fragments in the $^{236}$U$(n,f)$ reaction in one experiment as a function of neutron energy, from 0.3 to 500 MeV. To minimize systematic errors, the neutron-induced fission cross section of $^{236}$U was measured relative to that of $^{235}$U. At the same time, the experiment was set up in such a way that the angular distribution of fission fragments of ``reference'' $^{235}$U nuclei, which was widely studied, including by us (the corresponding references are given below), was additionally measured. In fact, our setup allowed us to carry out identical measurements for two different isotopes, $^{236}$U and $^{235}$U, due to the fact that identical conditions were created for them, namely: the same experiment geometry, the neutron flux and background conditions, the experimental setup and the data acquisition system, the processing procedure. The developed technique can be used to carry out similar measurements with other pairs of fissile isotopes.

The experiment was performed on the time-of-flight spectrometer of the neutron complex GNEIS \cite{Abrosimov_1985, Shcherbakov_2018} operating at the NRC ``Kurchatov Institute'' -- PNPI on the basis of the SC-1000 proton synchrocyclotron with a proton beam energy of 1~GeV. The fission fragments were registered using position-sensitive multiwire proportional counters (\mbox{MWPCs}) of low pressure \cite{Breskin_1982, Gagarski_2017}. The data acquisition system was organized on the basis of waveform digitizers, which allowed measurements to be made in a wide range of neutron energies with practically zero dead time.

The paper is organized as follows: after an introduction on the GNEIS facility and experimental apparatus in Sec.~I, the data analysis procedure is described in Sec.~II. Results are reported in Sec.~III together with a comparison with previous data and with current evaluations as well as with the theoretical description performed.

\section{Experiment}

\subsection{\label{s02-1} General information}

\begin{figure*}
\begin{center}
\includegraphics[scale=0.6]{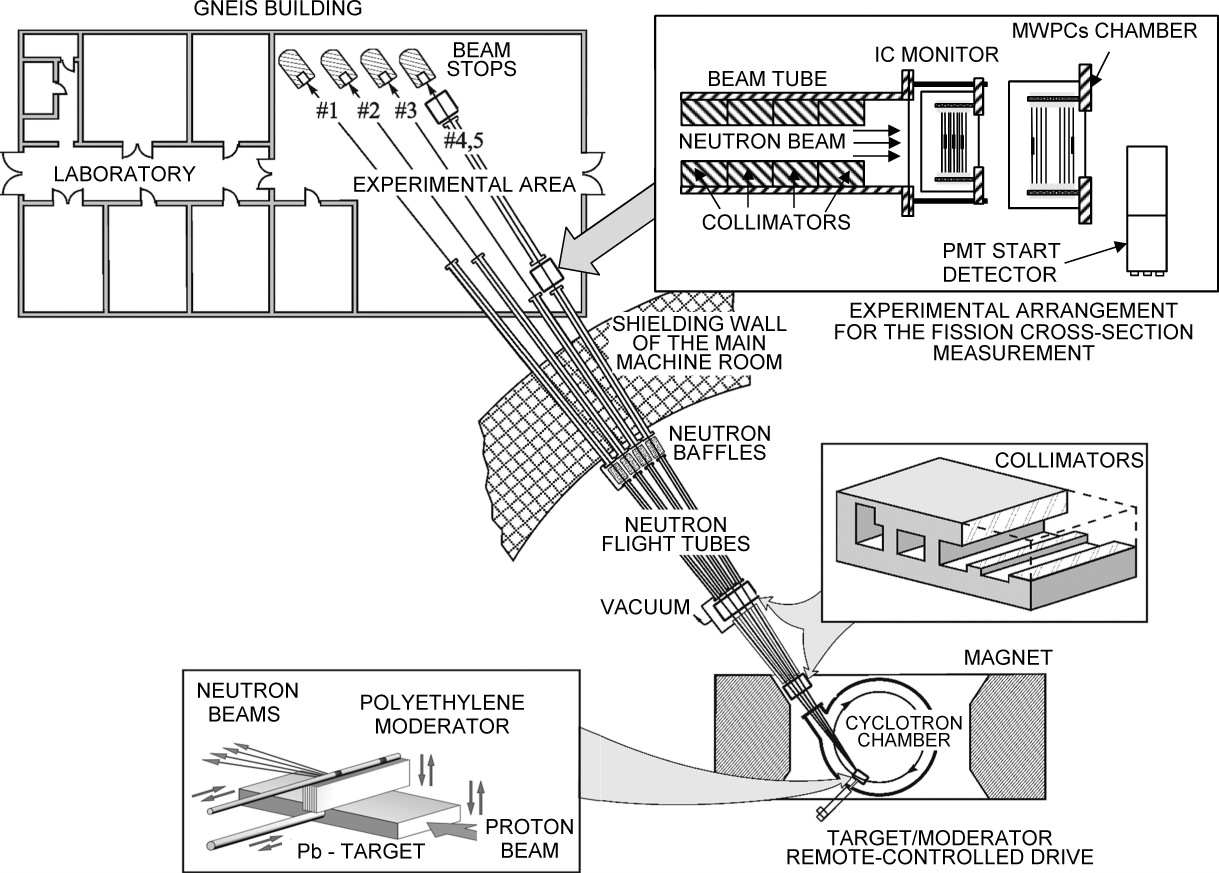}
\caption{General view of the GNEIS neutron complex. The layout of the experimental installation on the GNEIS spectrometer is shown in upper right corner.}\label{f01}
\end{center}
\end{figure*}

Measurement of the neutron-induced fission cross section of $^{236}$U and angular distributions of fragments in the reactions $^{236}$U$(n,f)$ and $^{235}$U$(n,f)$ were performed at the neutron complex GNEIS. The neutron complex GNEIS includes an intense pulsed neutron source ($\approx$ 10$^{14}$ neutrons/(sec 4$\pi)$) with the accelerator burst width of $\approx$ 10 ns and repetition frequency of $\approx$ 50 Hz, as well as a time-of-flight spectrometer with the bases up to 50 m long. These features of the GNEIS spectrometer make it possible to study the interaction of neutrons with atomic nuclei in a wide energy range from 0.01 eV to hundreds of MeV \cite{Shcherbakov_2018}. The general view of the neutron complex GNEIS and the location of the experimental setup are shown in Fig.~\ref{f01}.

The pulsed neutron source is the ``target + moderator'' system located in the vacuum chamber of the accelerator. A fast neutron pulse is formed as a result of the discharge of a proton beam onto a water-cooled lead target. The target and the polyethylene moderator located next to it can move independently in the radial and vertical directions both being remotely controlled from the accelerator console.

The shaping of neutron beams is carried out by a system of collimators located in the main hall of the SC-1000 and inside the evacuated neutron flight tubes. Neutron beams, equipped with cast-iron shutters, are led into a free-standing GNEIS building, consisting of an experimental hall and a laboratory and measurement section. Neutron beam stops are installed at the ends of the flight bases, approximately 50~m away from the neutron source.

The experimental setup for measuring fission cross sections and angular distributions consists of a set of \mbox{MWPCs} (see \mbox{MWPCs} chamber in Fig.~\ref{f01}), a fission ionization chamber with $^{238}$U targets for relative monitoring of the neutron flux (IC monitor in Fig.~\ref{f01}) and a photomultiplier tube located in a neutron beam to generate trigger signal of a neutron pulse (PMT - Start detector in Fig.~\ref{f01}).

\begin{figure}
\begin{center}
\includegraphics[scale=0.29]{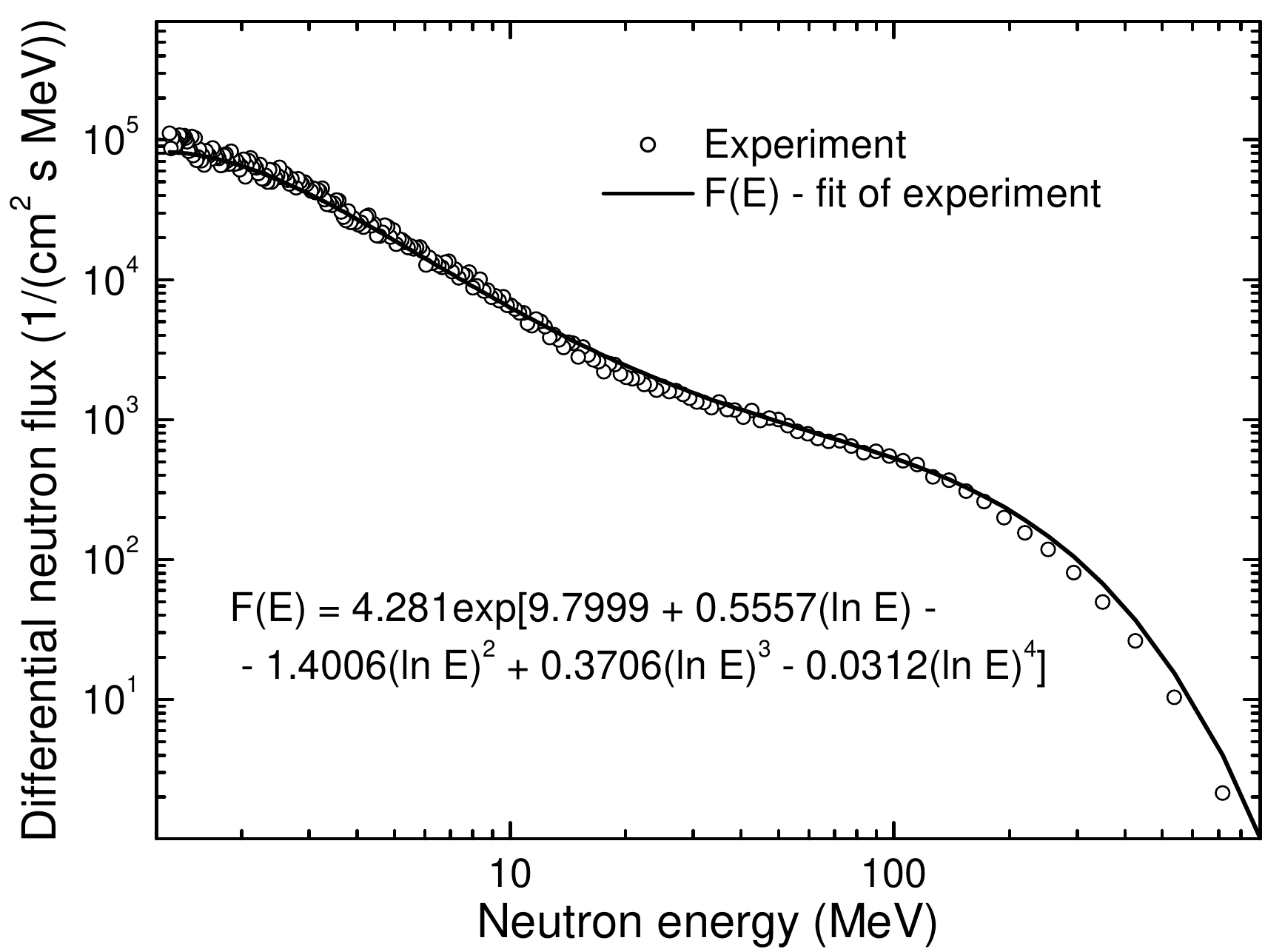}
\caption{Differential flux (energy spectrum) of neutrons measured for beam №~5 on a flight path of 36.5~m, which corresponds to integral flux of $4.2\times 10^5$~n/(cm$^2$ s) of neutrons with energies between 1 - 1000~MeV~\cite{Shcherbakov_2016}.}\label{f02}
\end{center}
\end{figure}

The time-of-flight spectrometer has 5 neutron beams; the axes of beams №~1-4 pass through the moderator, and the axis of beam №~5 --- through the lead target. As a result, the neutron spectra of beams №~1-4 are dominated by slow and resonant neutrons with energies below 100~keV, while the spectrum of beam №~5 is maximally ``rigid'' in the 0.3 – 500 MeV neutron energy range of interest \cite{Shcherbakov_2016}. Fig.~\ref{f02} shows the energy spectrum of neutrons of the beam №~5, measured using the IC monitor, and its fitting function $F(E)$. 

\begin{figure}
\begin{center}
\includegraphics[scale=0.25]{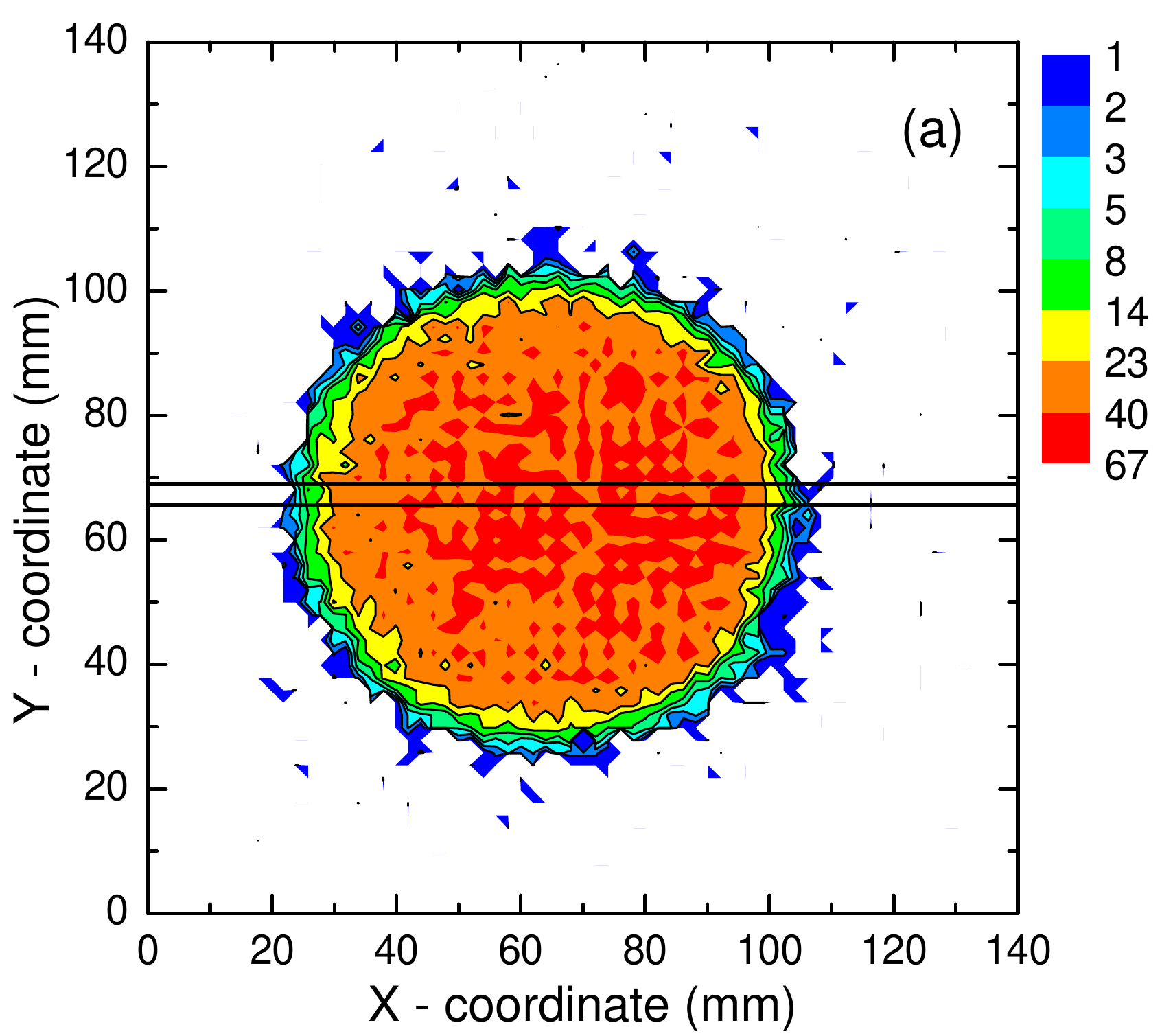}
\vspace{4mm}

\includegraphics[scale=0.29]{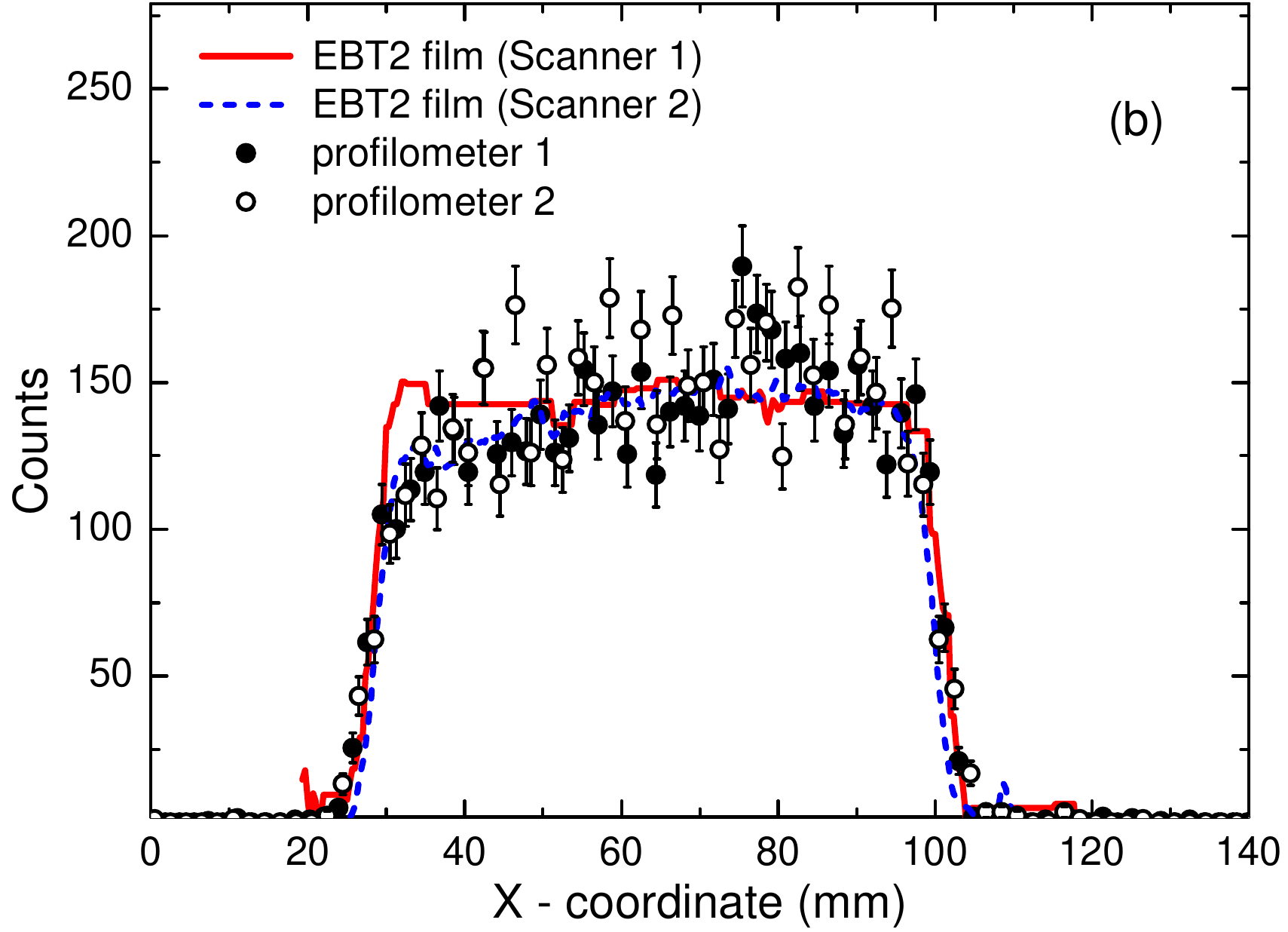}
\caption{(a) 2-D representation --- neutron beam profile measured using profilometer 1 \cite{Gagarski_2017, Shcherbakov_2016}. (b) Projection on the X-axis of the selected area of the 2-D representation; the diameter of the neutron beam collimator is 75 mm.}
\label{f03}
\end{center}
\end{figure}

Measurements were carried out on beam №~5, with a flight path length of 36.50(5) m. The cross section of the neutron beam was a circle with a diameter of 90 mm. The beam profile measured for a similar beam with a diameter of 75 mm using a GafChromic EBT2 radiometric film \cite{gafchromic} and two different neutron beam profilometers \cite{Gagarski_2017, Shcherbakov_2016} is shown in Fig.~\ref{f03}. In the same figure, the obtained distributions are compared. 

Note that the data obtained using the GafChromic EBT2 film are digitized using a scanner. Therefore, in order to estimate the error of the method, the film was scanned after irradiation with two scanners. Similarly, two different detectors based on \mbox{MWPC} have been used as two neutron beam profilometers. The first one, consisting of a single \mbox{MWPC}, was of the same type as that used in the work \cite{Shcherbakov_2016}. The second detector was an assembly of two \mbox{MWPCs} and a target with a fissile substance $^{238}$U (a converter of neutrons into fission fragments). This detector was similar to the one used to measure the angular distributions of fission fragments in \cite{Vorobyev_2015, Vorobyev_2016, Vorobyev_2018, Vorobyev_2019, Vorobyev_2020}. From the data presented in Fig.~\ref{f03}, it can be seen that the measurement results performed by different methods are in good agreement with each other.

It should be noted that the time between successive disposals of a proton beam onto a lead target is $\approx$ 20 ms, which at the flight path length of 36.5 m corresponds to the energy of recycled neutrons of less than 0.017 eV. To exclude such recycled neutrons, a 0.1 mm thick Cd filter was used, the transmission of which for neutrons with energies below 0.3 eV can be considered as equal to zero. 

Verification of the beam wiring system and its basic parameters such as spatial and energy distributions was carried out by comparing the measured dependencies with calculations performed by means of the Monte Carlo method \cite{Nakin_2019} using the Geant4 package \cite{Agostinelli_2003}. It was found that the simulation results reproduce the dependencies observed in the experiment.

\subsection{\label{s02-2} Targets}

\begin{figure}
\begin{center}
\includegraphics[scale=0.4]{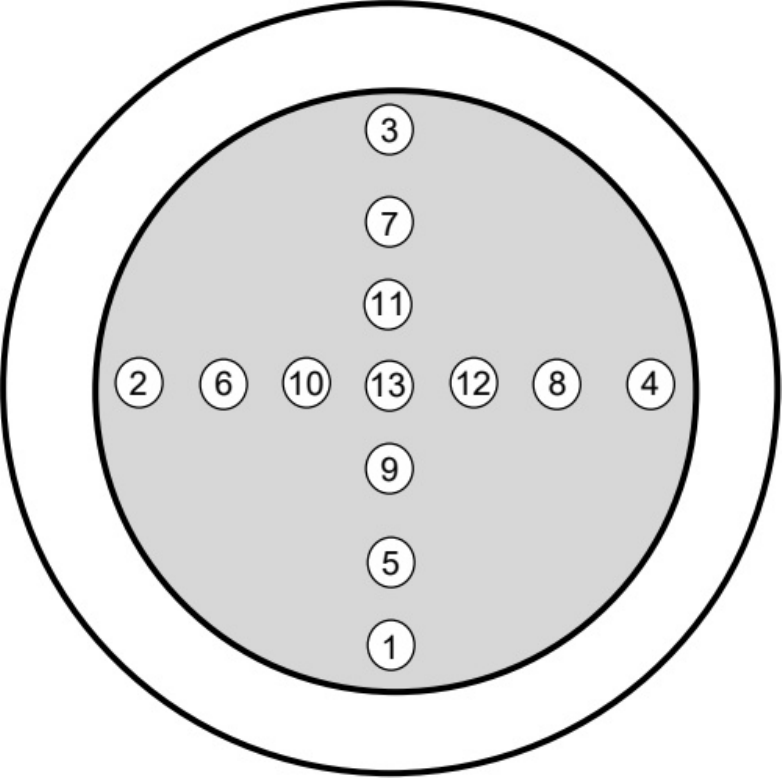}
\caption{Diagram of an aluminum substrate with an active layer applied (shaded area) --- target $^{236}$U. The letters mark the areas where measurements of $\alpha$-activity were made.}
\label{f04}
\end{center}
\end{figure}

The targets with the $^{236}$U and $^{235}$U were manufactured at JSC ``V.G. Khlopin Radium Institute'' (St. Petersburg, Russia) by the ``painting'' technique, in which several layers of the compound of the studied isotope (as part of the chemical compound U$_3$O$_8$) are applied to aluminum foil 0.1 mm thick, each one was annealed at high temperature. The number of layers applied was determined on the basis of ensuring the uniformity of the active layer $\approx$ 10\%. We present here the information about parameters of the targets (Table~\ref{t1}), isotopic composition of the targets (Table~\ref{t2}), uniformity of the active layer (Fig.~\ref{f04} and Table~\ref{t3}), taken from the manufacturer's certificate. It can be seen from the Table~\ref{t1} that the shape and size of the active layers were different. The isotopic composition of the targets shown in the Table~\ref{t2} was determined by mass spectrometry. For completeness, half-lives of the corresponding isotopes taken from the ENSDF database \cite{LCN} have also been added. The uniformity of the active layer was investigated by scanning the $\alpha$-activity of the target area using semiconductor detectors with a small solid angle. For example, Fig.~\ref{f04} schematically shows the target $^{236}$U and the points at which the $\alpha$-activity measurements were carried out. The measurement results are presented in the Table~\ref{t3} as the deviation of $\alpha$-counts at the point from the value obtained averaging over all selected points, statistical accuracy of measurements was 1.2 - 1.5\%.

\begin{table}
\caption{\label{t1} Target parameters}
\begin{center}
\begin{tabular}{|c|c|c|}
\hline
Main isotope & $^{235}$U & $^{236}$U \\
\hline
Thickness of active layer ($\mu$g/cm$^2$) & 203(11) & 317(16)  \\
\hline
Sizes of active layer (mm) & $50\times 100$ & $\oslash~82$ \\
\hline
Target mass (mg) & 10.15(51) & 16.70(83) \\
\hline
\end{tabular}
\end{center}
\end{table}

\begin{table}
\caption{\label{t2} Target composition}
\begin{center}
\begin{tabular}{|c|c|c|c|}
\hline
  &  & $^{235}$U & $^{236}$U \\
\hline
Isotope  & $T_{1/2}$ ($yr$) & \multicolumn{2}{|c|}{Mass percentage (\%)} \\
\hline
$^{234}$U & 2.455(6)$\times$ 10$^5$ & 0.0020(5) & \textless 0.00001 \\
\hline
$^{235}$U & 7.04(1)$\times$ 10$^8$ & 99.9920(10) & 0.0043(1) \\
\hline
$^{236}$U & 2.342(4)$\times$ 10$^7$ & 0.0040(5) & 99.9730(2) \\
\hline
$^{238}$U & 4.468(6)$\times$ 10$^9$ & 0.0020(5) & 0.0227(2) \\
\hline
\end{tabular}
\end{center}
\end{table}

\begin{table}
\caption{\label{t3} Non-uniformity of $^{236}$U target}
\begin{center}
\begin{tabular}{|c|c|c|c|c|c|}
\hline
Point & \parbox{1.5cm}{Deviation (\%)} & Point & \parbox{1.5cm}{Deviation (\%)} & Point & \parbox{1.5cm}{Deviation (\%)} \\
\hline
1 & -17.5 & 5 & 2.4 & 9 & -5.5 \\
\hline
2 & -4.8 & 6 & 6.3 & 10 & 1.1 \\
\hline
3 & -2.6 & 7 & 11.9 & 11 & -4.7 \\
\hline
4 & 0.0 & 8 & 7.8 & 12 & 0.7 \\
\hline
 &  &  &  & 13 & -2.1 \\
\hline
\end{tabular}
\end{center}
\end{table}

To ensure the identity of the conditions when measuring the fission cross sections on the $^{235}$U and $^{236}$U targets, we used ``masks'' made of 0.1 mm thick aluminum foil with a hole in the form of a circle with a diameter of 48.0(1) mm. These masks were put on the targets from the side of the active layer. Thus, round areas of the same size of the active substance involved to the measurements were selected. To determine the masses of the substance in these areas we measured the $\alpha$-activities of the masked targets using semiconductor detectors. The masses were found from the measured activities. In this case, the data on the isotopic composition of the targets and the half-lives of the corresponding isotopes presented in the Table~\ref{t2} were used. The masses were found to be 3.480(48) mg and 5.796(44) mg with relative statistical uncertainty 0.9\% and 0.2\% for the $^{235}$U and $^{236}$U nuclei, respectively. The average substance thicknesses in the selected circles found from these masses  are 192(3)~$\mu$g/cm$^2$ and 320(2)~$\mu$g/cm$^2$ for the targets of $^{235}$U and $^{236}$U, respectively. These values, within the limits of error, agree with the data on thickness given in the Table~\ref{t1}.

\subsection{\label{s02-3} Measurement procedure}

\begin{figure}
\begin{center}
\includegraphics[scale=0.62]{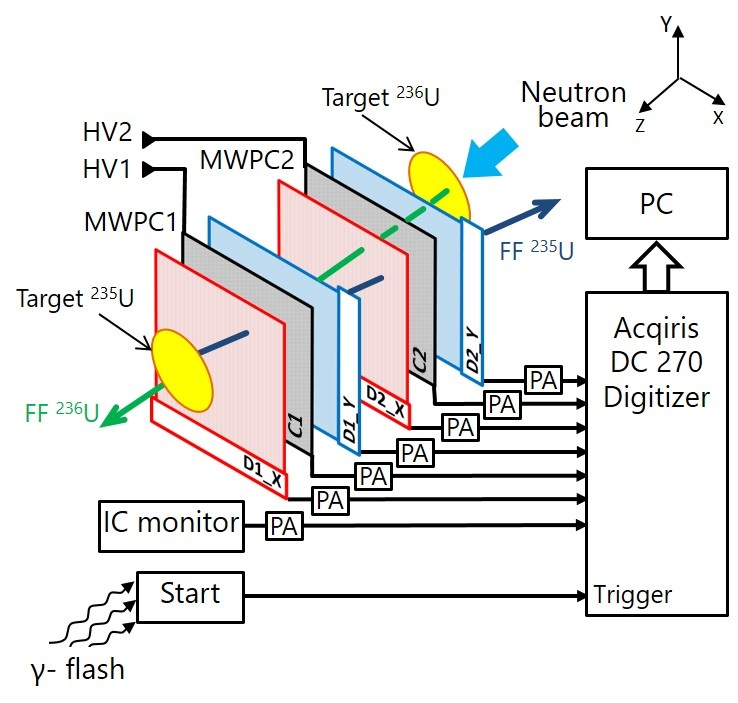}
\caption{General view of the experimental setup and data acquisition system: Start -- START detector; IC monitor -- the fission ionization chamber with $^{238}$U targets; PA -- preamplifier; HV1, HV2 -- high–voltage power sources; anodes D1\underline{\phantom{0}}X, D2\underline{\phantom{0}}X -- detectors 1, 2 ($X$ axis); anodes D1\underline{\phantom{0}}Y, D2\underline{\phantom{0}}Y -- detectors 1, 2 ($Y$ axis); C1, C2 are the cathodes of \mbox{MWPC1} and \mbox{MWPC2}, respectively.}
\label{f05}
\end{center}
\end{figure}

Measurements of the fission cross sections of $^{236}$U in the region of neutron energies 0.3 - 500 MeV were carried out relative to the fission cross section of $^{235}$U, which is known with high accuracy in this region of neutron energies and is recommended as fission cross section standard \cite{Marcinkevicius_2015, Carlson_2018}. A general view of the experimental setup and the data acquisition system is shown in Fig.~\ref{f05}. This setup is a modified version of the setup previously used for measuring the angular distributions of fission fragments \cite{Vorobyev_2015, Vorobyev_2016, Vorobyev_2018, Vorobyev_2019, Vorobyev_2020, Vorobyev_2020-2}. The main features of the new installation are discussed below.

During the measurements, fission fragments emitted from the target of the studied isotope $^{236}$U and from the target with the reference isotope $^{235}$U were recorded in the same measuring session by an assembly of two position-sensitive detectors (\mbox{MWPCs}), which were placed in a cylindrical chamber filled with the working gas isobutane C$_4$H$_{10}$ at a pressure of 8 mbar. The assembly of the \mbox{MWPCs} and targets is located perpendicular to the neutron beam (which can be taken as the Z axis).

Each of the two \mbox{MWPCs} consists of three wire electrodes: two anodes and one cathode. The planes of the anodes D\underline{\phantom{0}}X and D\underline{\phantom{0}}Y are gold-plated tungsten wires with a diameter of 25~$\mu$m, parallel to axes X and Y, respectively. The distance between the wires is 1 mm and the anode-cathode gaps are 3.4 mm. The cathodes C are grids of the same wires. The size of the sensitive area of each \mbox{MWPC} was $140\times 140$ mm$^2$. The total efficiency of the detector (assembly of two \mbox{MWPCs}), equal to $\approx$ 45\%, is determined by the solid angle of registration and the transparency of 3 anodes and 2 cathodes through which the fragment must fly before it is registered.

Targets with the studied and reference isotope were placed on both sides relative to the assembly of \mbox{MWPCs} parallel to the planes of the electrodes. The distances from each of the targets to the cathodes of two consecutive counters were equal to 6.8 mm and 37 mm. The maximum registration angle $\theta$ relative to the normal to the plane of the \mbox{MWPCs} electrodes was $\theta_{max}\simeq 71^{\circ}$ ($\cos\theta > 0.325$).

Signals from 2 anodes and cathode of each \mbox{MWPC}, as well as a signal from the IC monitor and a signal from the START detector were fed through fast preamps to 8 inputs of signal shape digitizers (Acqiris DC-270, 8-bit resolution, sampling rate 500 MSamples/s). The signal shape digitizers were launched synchronously with each disposal of a proton beam onto a lead target using signals from the START detector, the acquisition time window for digitizing signals across all 8 inputs of the converter was 8~$\mu$s, which corresponds to incident neutron energy from 1 GeV down to $\approx$ 110 keV. Further, the signals were read into the computer and stored on the hard disk for operational control of the received information and subsequent off-line processing.

The signals from cathodes C1 (of the \mbox{MWPC1}) and C2 (of the \mbox{MWPC2}) give time marks of the passage of fission fragments through them. The signals from the anodes D1\underline{\phantom{0}}X, D1\underline{\phantom{0}}Y (of the \mbox{MWPC1}) and from the anodes D2\underline{\phantom{0}}X, D2\underline{\phantom{0}}Y (of the \mbox{MWPC2}) are used to determine the coordinates of the detected particle (x1, y1) and (x2, y2), respectively. Each anode consists of 140 wires connected in pairs with 70 taps of the corresponding delay lines (impedance 50 ohms, delay 2 ns per tap). One end of each delay line is grounded, and the time signals taken from the other end of the delay line carry information about the corresponding coordinate. The coordinates depend linearly on the time difference between the signals from the cathodes and the corresponding anodes.

\begin{figure}
\begin{center}
\includegraphics[scale=0.35]{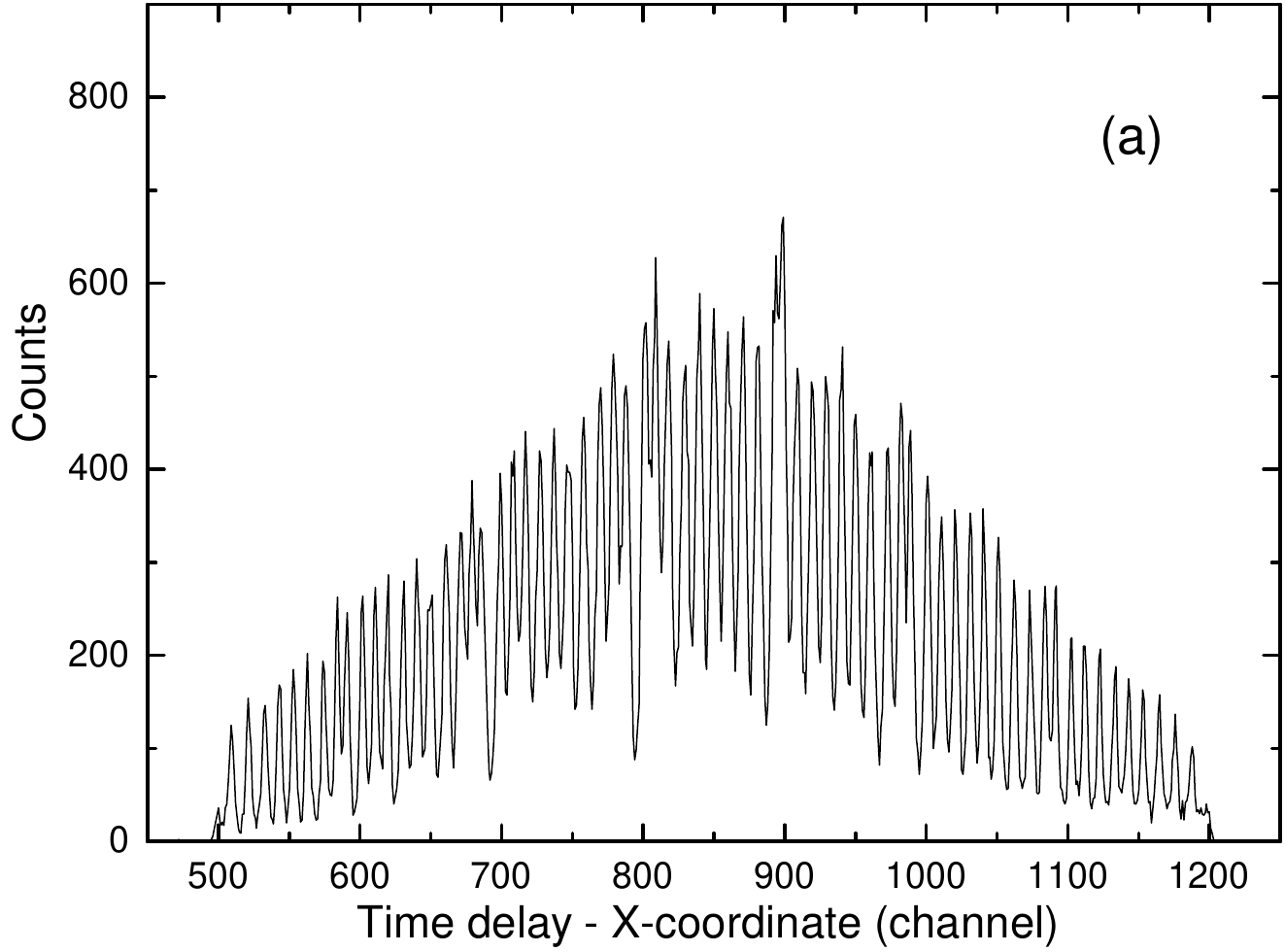}
\includegraphics[scale=0.35]{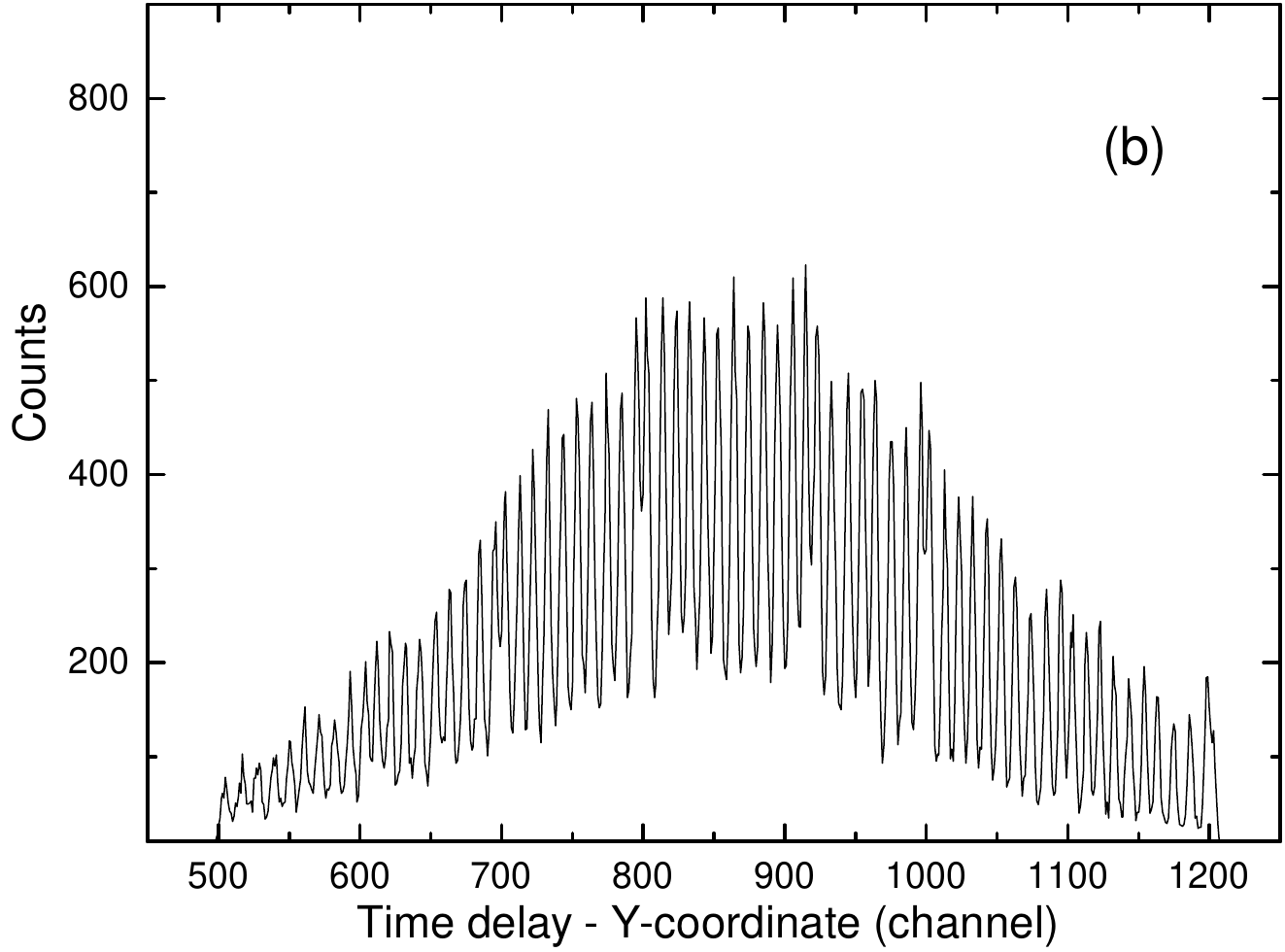}
\caption{Time distribution of $^{235}$U fission fragments from the anodes of \mbox{MWPC2}: (a) --- for the anode D2\underline{\phantom{0}}X; (b) --- for the anode D2\underline{\phantom{0}}Y. The channel width is 0.2 ns, which corresponds to 0.2 mm.}
\label{f06}
\end{center}
\end{figure}

As can be seen from Fig.~\ref{f06}, where, for example, the time distributions of signals from $^{235}$U fission fragments from the anodes D2\underline{\phantom{0}}X, D2\underline{\phantom{0}}Y of the \mbox{MWPC2} are presented, it is easy to identify peaks corresponding to events registered by each of the 70 pairs of wires (step 2 mm). From these distributions, the spatial resolution in x and y coordinates was found to be 2 mm. The effect of the obtained resolution on the measured angular distributions of fission fragments and on the anisotropy of these distributions was taken into account by the Monte Carlo method (see Secs.~\ref{s03-1},~\ref{s03-3}).

The cosine of the angle $\theta$ between the fission axis and the normal to the \mbox{MWPCs} plane was calculated using the following expression:
\begin{equation}
\label{1}
\cos\theta=\frac{d}{\sqrt{(x1-x2)^2+(y1-y2)^2+d^2}}\,
\end{equation}
where $d$ is the distance between cathodes C1 and C2.

It is worth noting that the main advantages of position-sensitive \mbox{MWPC}, which make them almost an ideal tool for registering fission fragments, are the following: excellent time characteristics --- the full width of the signal at half the height is $\approx$ 16 ns, the registration efficiency is close to unity, high transparency and low energy losses inside the detector, a large registration area, high counting speed, good spatial resolution, insensitivity to $\gamma$-flash, stable operation. This made it possible to reliably isolate fission events against the background of such reactions as the natural $\alpha$-activity of the target nucleus and various reactions $(n,x)$ on the target nuclei, substrate and other structural materials of the detector. Taking into account the fact that the data acquisition system was organized on the basis of signal form digitizers, it also allowed measurements to be made in the range of 0.3 - 500 MeV neutron energies with almost zero dead time.

\begin{figure}
\begin{center}
\includegraphics[scale=0.43]{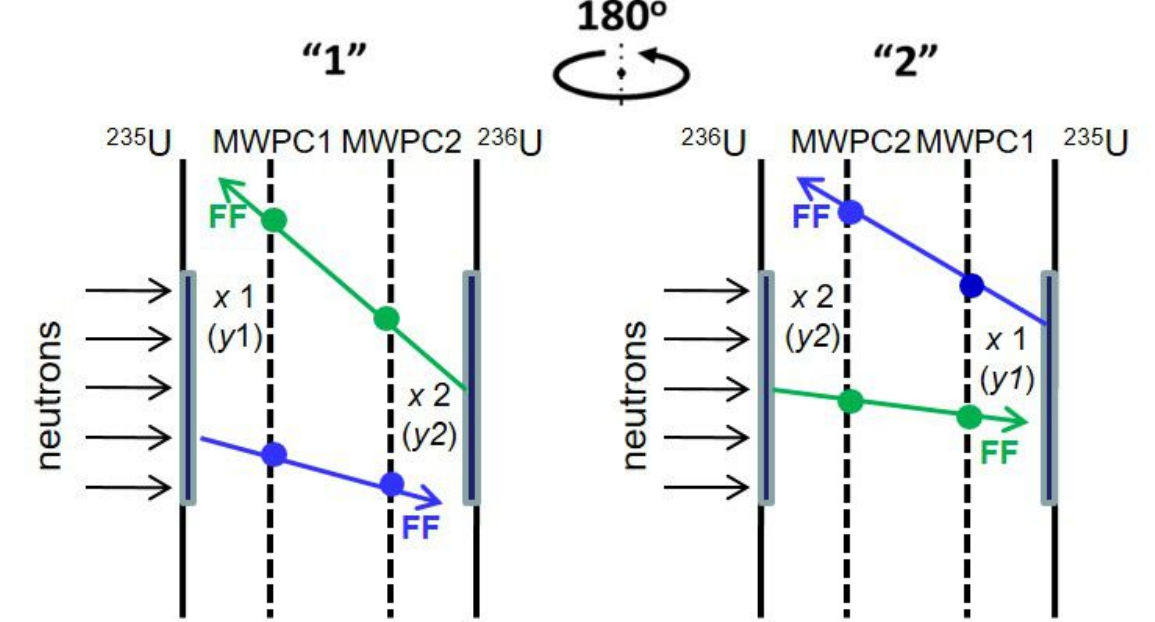}
\caption{Diagram of the relative position of the fission fragment detectors for two orientations of the installation relative to the incident neutron beam: ``1'' --- orientation 1; ``2'' --- orientation 2; $^{235}$U and $^{236}$U --- aluminum substrate with a layer of $^{235}$U and $^{236}$U; FF --- fission fragment; x1 and y1 --- coordinates of the particle on the anodes D1\underline{\phantom{0}}X, D1\underline{\phantom{0}}Y of counter 1; x2 and y2 are the coordinates of the particle at the anode D2\underline{\phantom{0}}X and D2\underline{\phantom{0}}Y of counter 2.}
\label{f07}
\end{center}
\end{figure}

As is known, the angular distribution of fragments in the laboratory coordinate system is distorted due to the fact that the neutron causing fission transmits an impulse to the fissionable nucleus. To account for this effect, measurements of the fission cross sections and angular distributions of fragments were performed for two orientations of the installation relative to the incident neutron beam: 1 --- the beam direction and the longitudinal component of the fragment pulse from the target under study $^{236}$U are oppositely directed and 2 --- the beam direction and the longitudinal component of the fragment pulse from the target under study coincide. The orientation change was achieved by rotating the installation 180$^{\circ}$ around an axis perpendicular to the direction of neutron motion in the beam and passing through the center of the installation. The scheme of the mutual arrangement of the detectors of fission fragments \mbox{MWPC1} and \mbox{MWPC2} is shown in Fig.~\ref{f07}.

The kinetic energy $E$ of the neutrons causing fission was determined by the time-of-flight method, which consists in the fact that by measuring the time of flight $t$ by a neutron of a fixed distance $L$, it is possible to obtain the value of the neutron velocity $v=L/t$. Thus, the kinetic energy is given by:
\begin{equation}
\label{2}
E=mc^2\left(\frac{1}{\sqrt{1-\left(L/(ct)\right)^2}}-1\right)\,,
\end{equation}
where $m$ is the mass of the neutron, $L$ is the flight base (the distance from the neutron source of the GNEIS spectrometer located in the vacuum chamber of the SC–1000 synchrocyclotron to the active layer with a fissile isotope), $c$ is the speed of light.

During measurements on the GNEIS spectrometer, the mark that a proton beam was dropped onto the lead target of the GNEIS neutron source is a signal from the START detector that registers $\gamma$-quanta and neutrons emitted from this target. At the signal from the START detector, the measuring system is started. As a result of the analysis of waveforms obtained from digitizers, the signals corresponding to the fission fragments are isolated and their time and amplitude characteristics are determined, from which the studied dependencies on the neutron energy are formed. The procedure for preliminary processing, introduction of necessary corrections and data analysis is presented below.

\section{Data analysis}

\subsection{\label{s03-1} Selection of fission events}

\begin{figure}
\begin{center}
\includegraphics[scale=0.37]{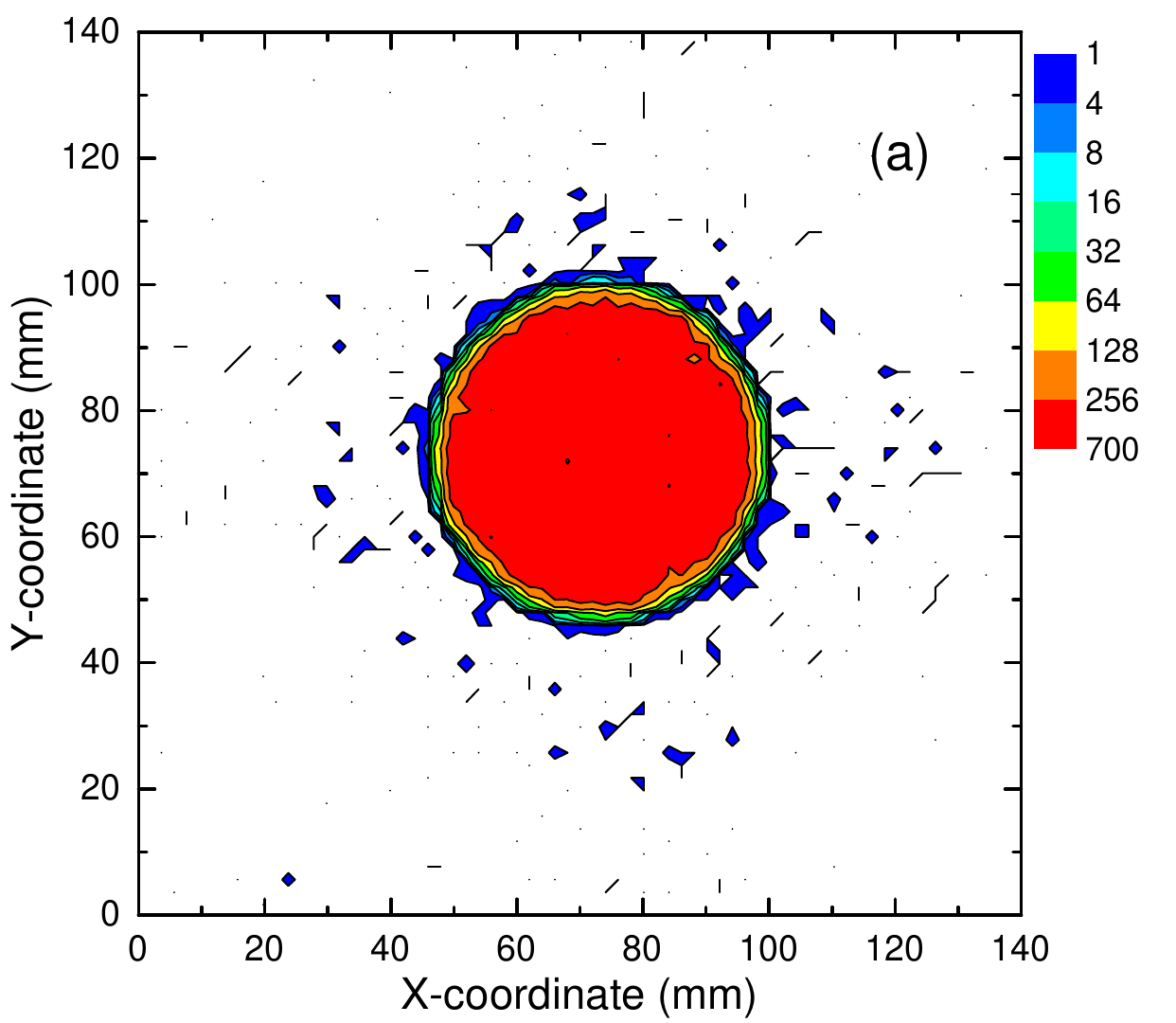}
\includegraphics[scale=0.37]{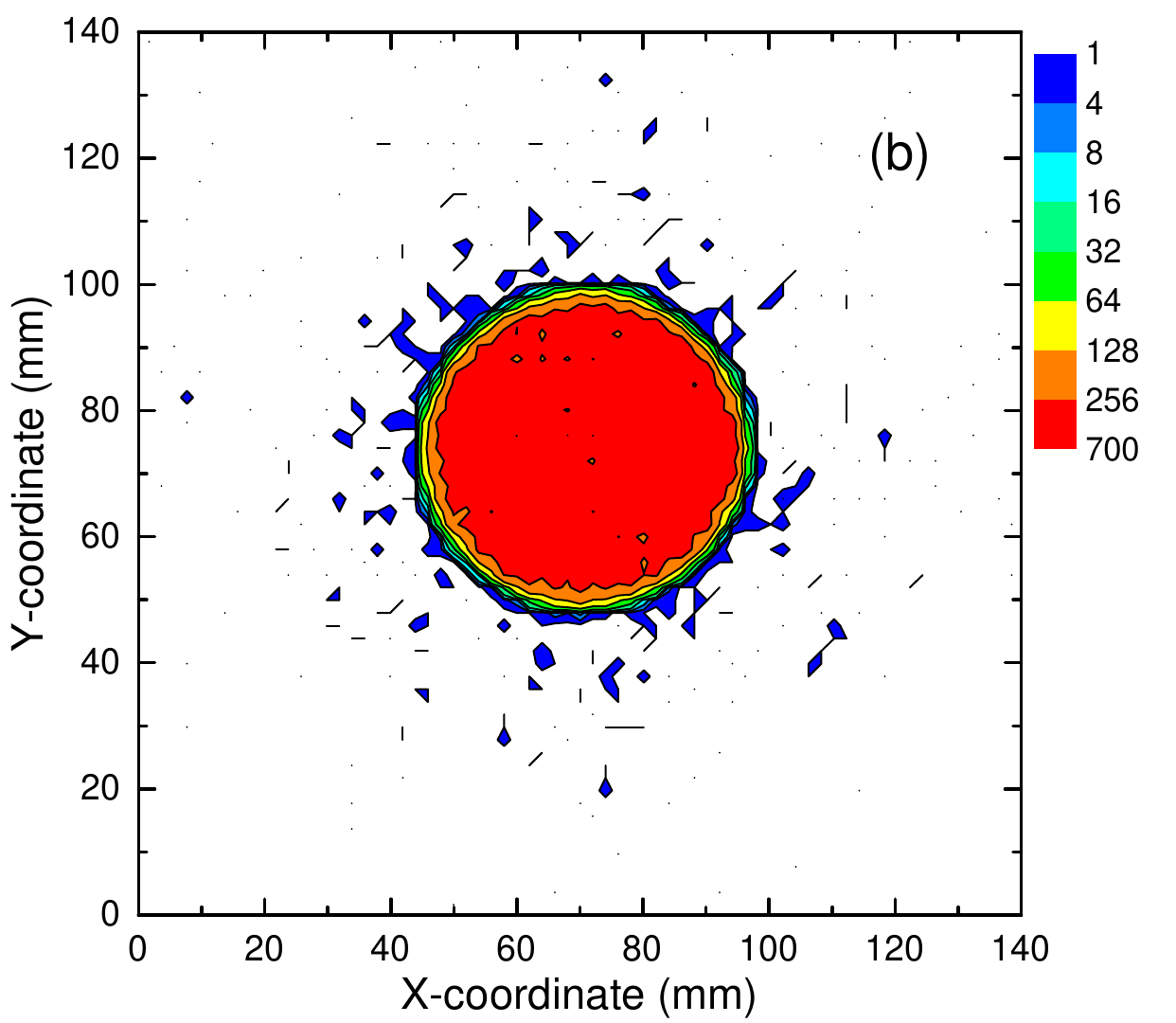}
\caption{Spatial distribution of selected fission events from target $^{236}$U (a) and $^{235}$U (b).}
\label{f08}
\end{center}
\end{figure}

In this paper, the procedure for separating fission events was organized in a manner similar to that described quite fully in \cite{Vorobyev_2016, Vorobyev_2018}. The essential point is that the implemented method of event selection allows not only to separate the fission events from background events, but also to select those fission events that correspond to fragments, both stopped on the wires of the \mbox{MWPCs} electrodes and flying through the counter without colliding with the wires. The fission events corresponding to the latter case were almost completely separated from background reactions induced by neutrons in the target backing and on other detector materials. Figure~\ref{f08} shows the spatial distribution of the selected fission events from targets $^{236}$U and $^{235}$U for one measurement series. It can be seen that the measured distributions have the shape of a circle with a diameter corresponding to the diameter of the active layer --- 48 mm. Since the diameter of the neutron beam during measurements was 90 mm, the absence of events outside the boundaries of the active layer of the target demonstrates the reliability of the procedure for separating fission events. 

\begin{figure*}
\begin{center}
\includegraphics[scale=0.35]{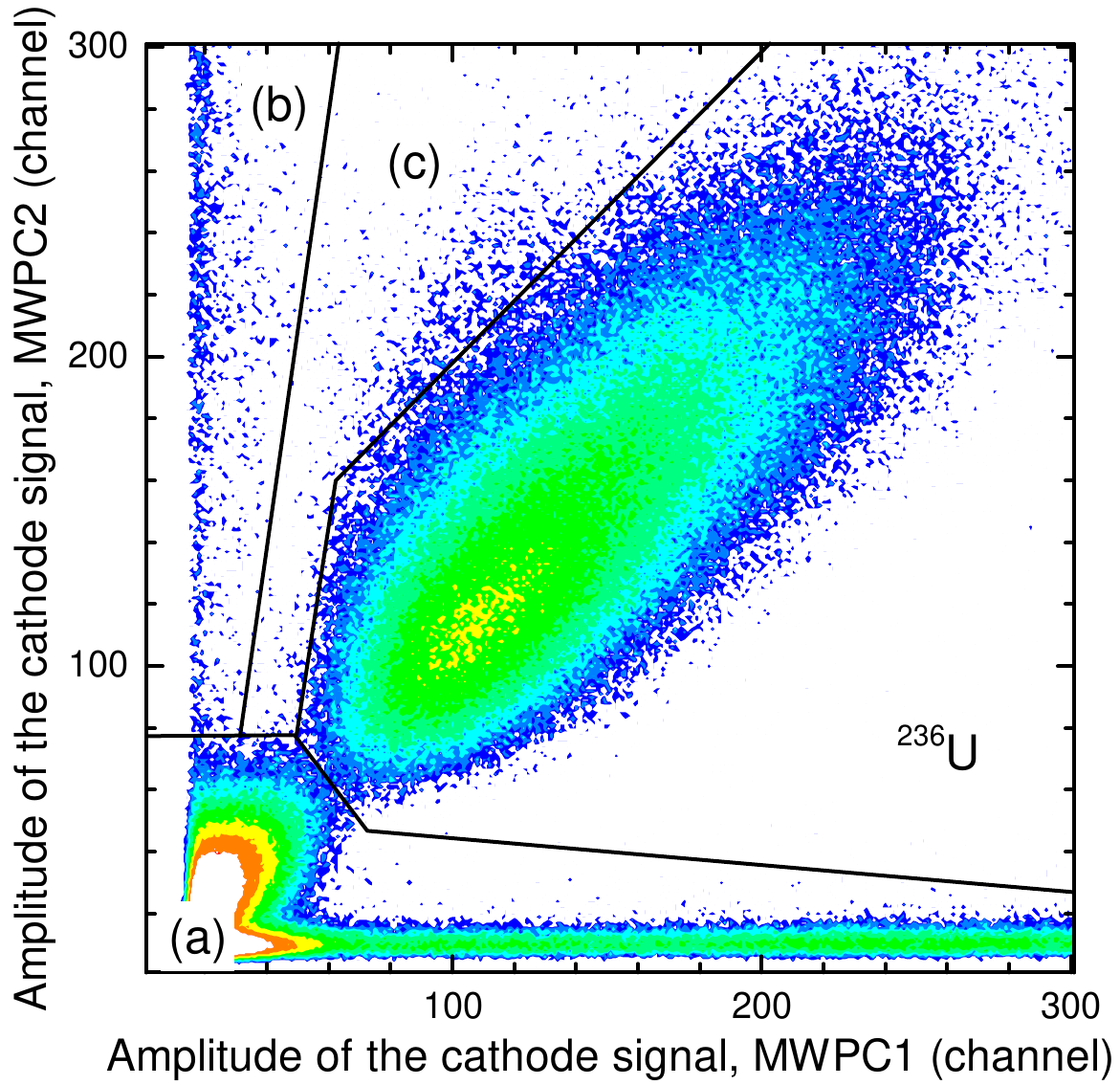}
\includegraphics[scale=0.35]{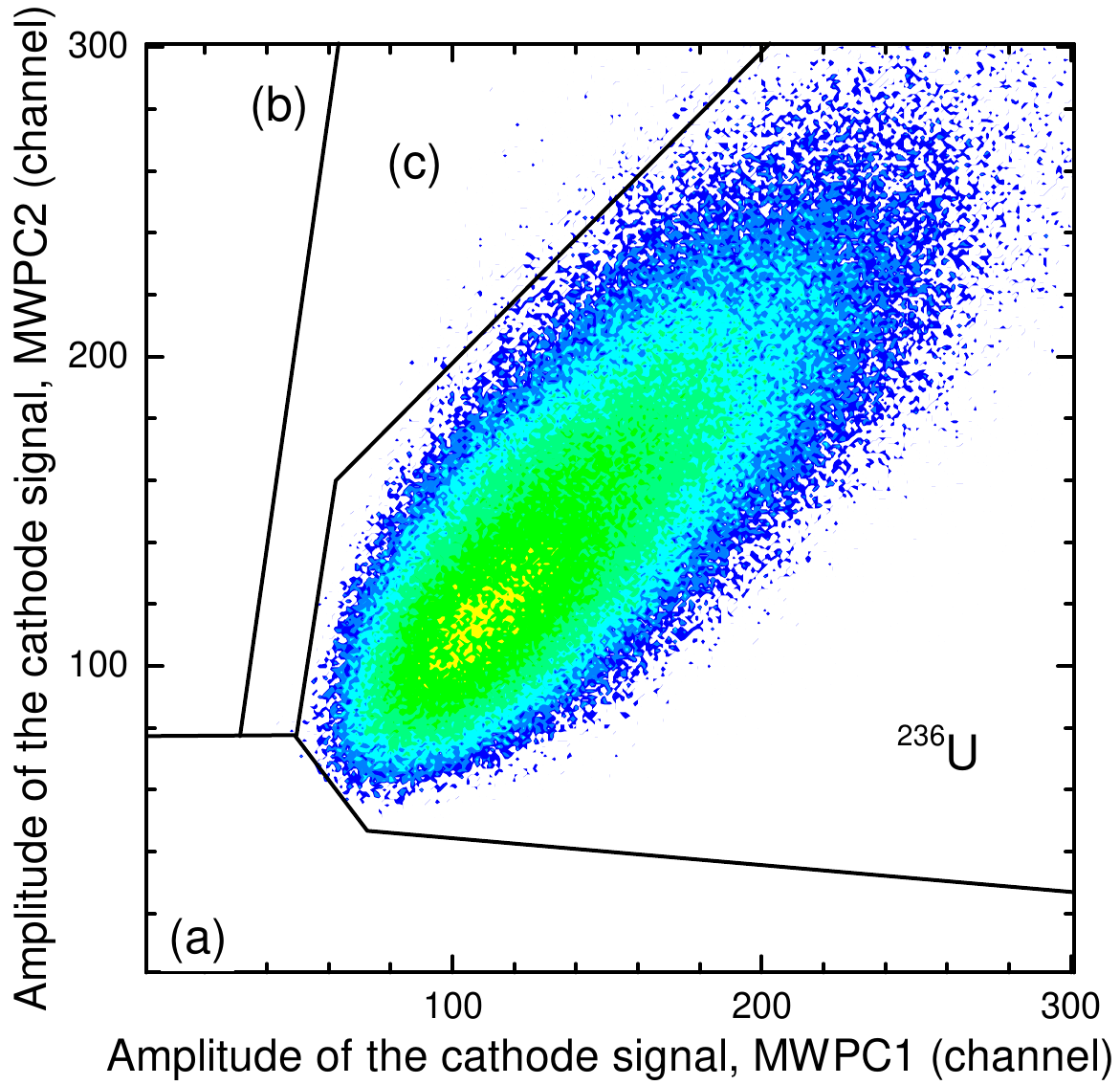}
\vspace{1mm}

\includegraphics[scale=0.35]{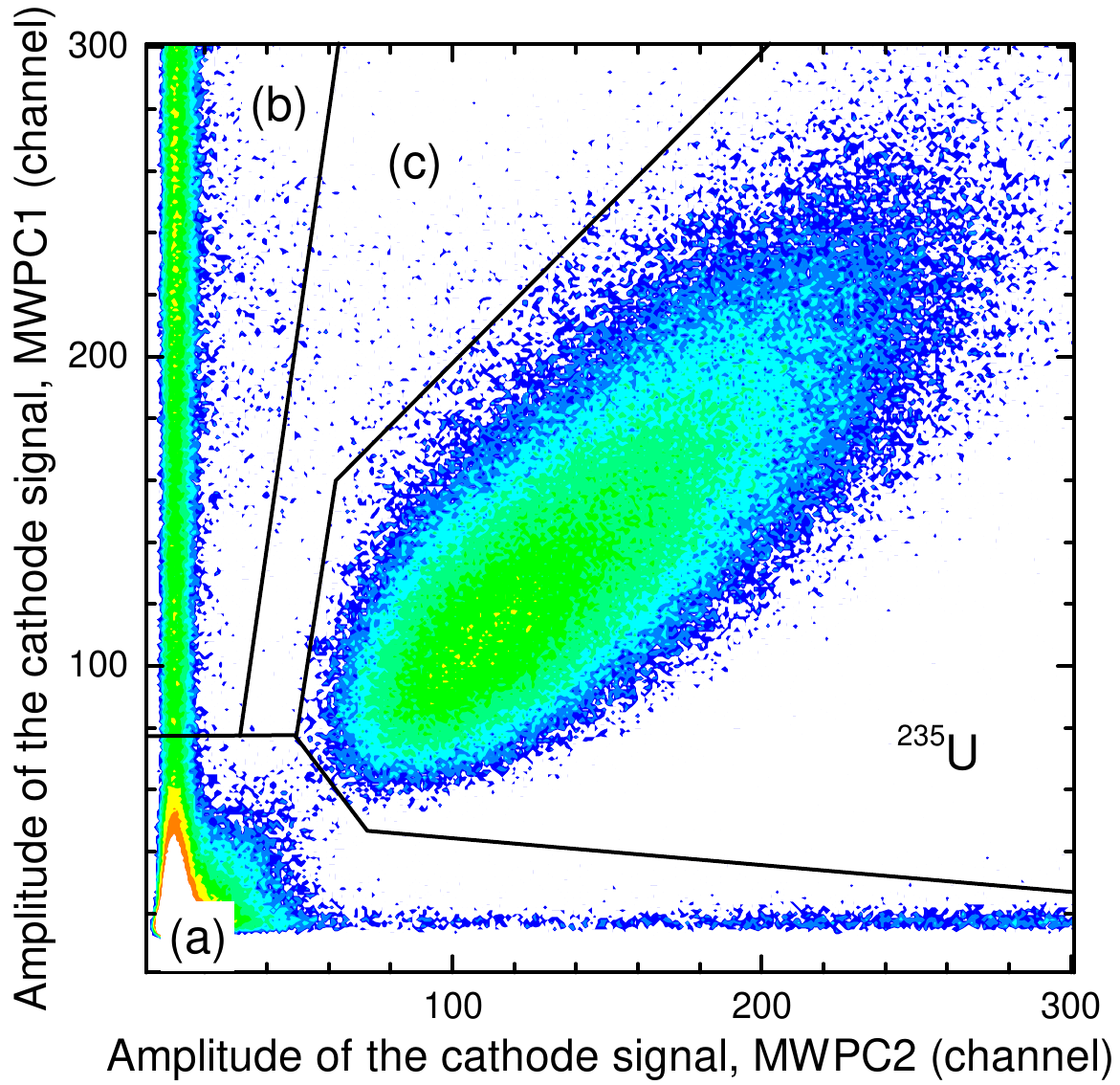}
\includegraphics[scale=0.35]{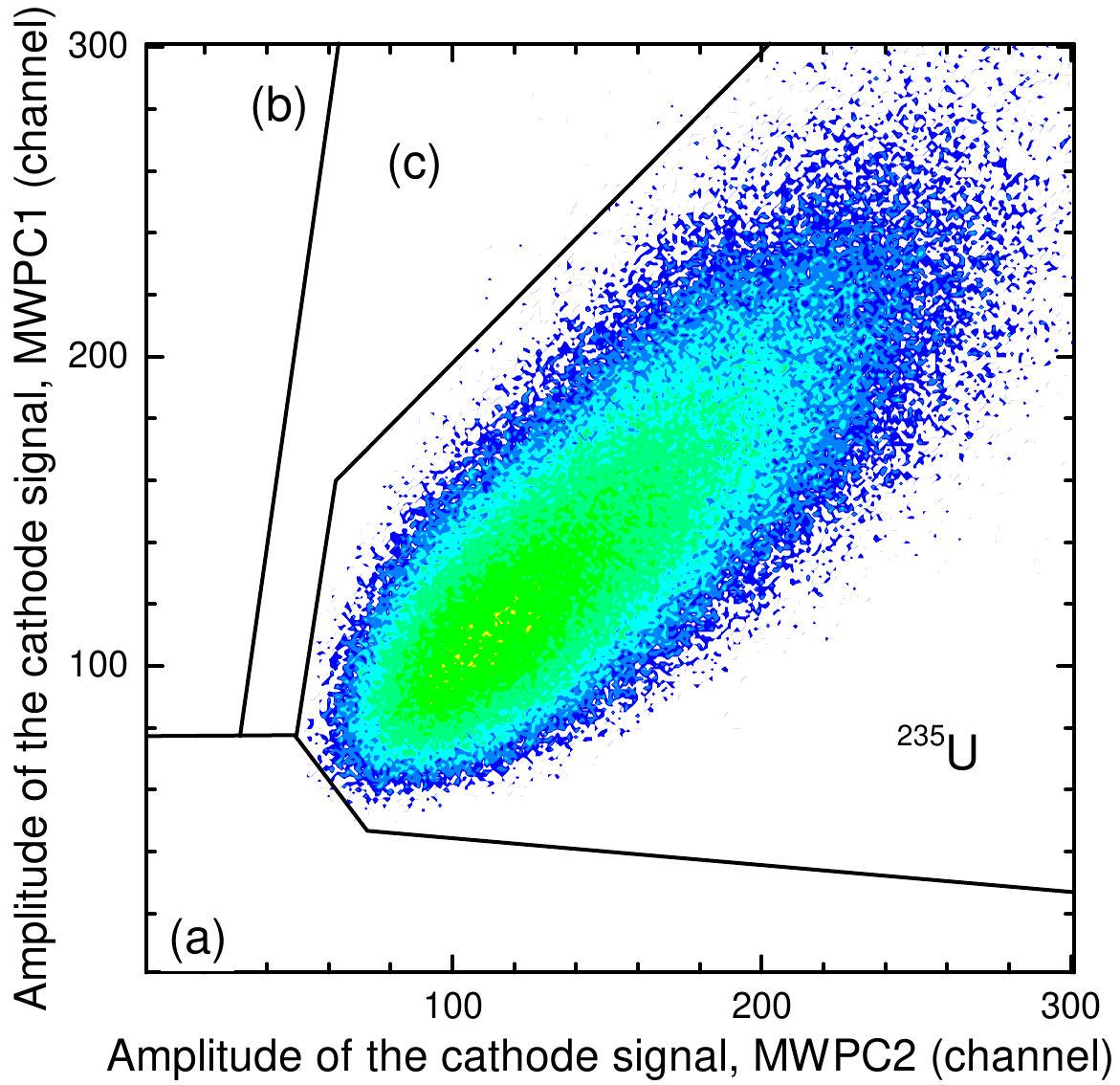}
\caption{On the left, a two–dimensional representation of the dependence between the amplitudes of cathode signals from two \mbox{MWPCs} in an experiment with $^{236}$U and $^{235}$U targets: (a) is the region of non-fission products of neutron-induced reactions and events of natural $\alpha$-activity; (b) and (c) are fission events corresponding to the case when fission fragments are registered only by one of the \mbox{MWPCs} due to absorption on the wires of the other. The right part of the figure shows the ``useful'' fission events (these results were obtained for orientation 2 of the installation relative to the direction of the neutron flux).}
\label{f09}
\end{center}
\end{figure*}

The two-dimensional distributions of recorded events before and after the separation of ``useful'' fission events (during registration of which the fragments do not stop at the wire electrodes of the \mbox{MWPCs}), shown in Fig.~\ref{f09}, demonstrate the high efficiency of the procedure used. For further analysis, only ``useful'' events are used. 
It should be noted that in this experiment, compared to the previously performed \cite{Gagarski_2017, Vorobyev_2015, Vorobyev_2016, Vorobyev_2018, Vorobyev_2019, Vorobyev_2020, Vorobyev_2020-2}, the distance between anodes D1\underline{\phantom{0}}Y and D2\underline{\phantom{0}}X increased from 3.4 mm to 23.4 mm. As a result, it was possible to completely avoid the distortion of the measured angular distributions of fragments due to the mutual influence of signals from the anodes of two neighboring \mbox{MWPCs} (the so-called ``cross-talk'' effect).
In this case, the efficiency of detecting fragments is completely determined by the geometric transparency of the wire electrodes and, therefore, when calculating the efficiency, it is sufficient to take into account the following detector parameters: the structure of the wire electrodes, the distance between the \mbox{MWPCs} and the targets, the size of the electrodes and the distance between them.

\begin{figure}
\begin{center}
\includegraphics[scale=0.28]{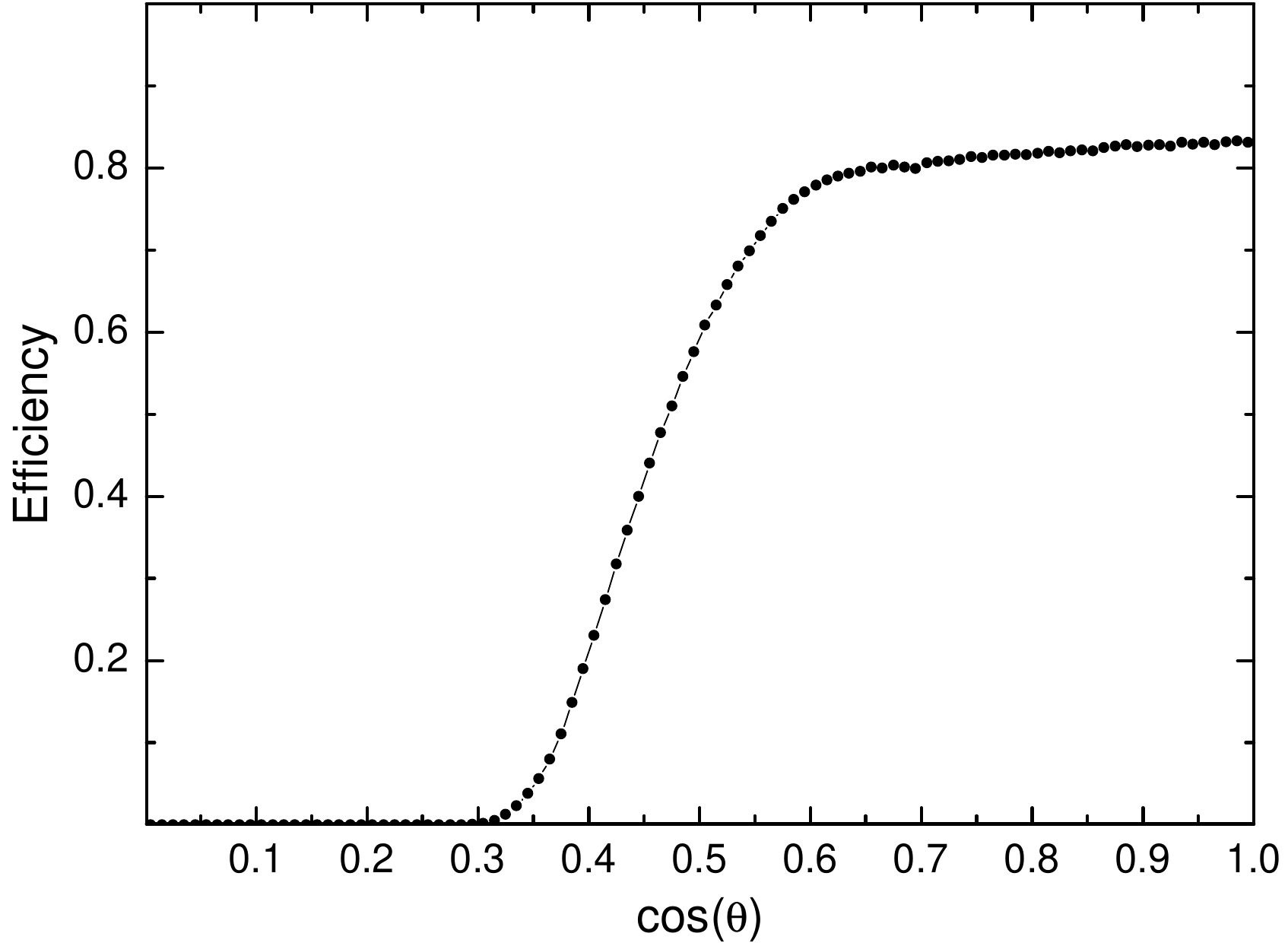}
\caption{Dependence of the efficiency of registration of fission fragments, $\varepsilon$, on the cosine of the departure angle $\theta$ relative to the normal to the target plane (calculation by the Monte Carlo method).}
\label{f10}
\end{center}
\end{figure}

The calculation of the efficiency of registration of fission fragments $\varepsilon$ by an assembly of two position-sensitive detectors was carried out using the Monte Carlo method, in which, in addition to the geometry of the \mbox{MWPCs}, the following features related to the measurement procedure were also taken into account: the dimensions of the target and the neutron beam, the spatial resolution of the \mbox{MWPCs}. The result of the efficiency $\varepsilon(\theta)$ calculation is shown in Fig.~\ref{f10}. The efficiency uncertainty associated with each parameter used in the Monte Carlo calculation was defined as the difference between the calculation result obtained by varying that parameter within its known uncertainty and the result of the primary efficiency calculation. These uncertainties were then considered independently of each other. So, it was found that the uncertainty due to the finite angular resolution reaches its maximum value of 0.3\% in the region of angles close to $0^{\circ}$ and decreases with increasing angle $\theta$. The total efficiency uncertainty due to geometric uncertainties was obtained by quadratic summing the uncertainties of the parameters mentioned above and was less than 0.5\% in the entire range of angles, except for angles close to $\theta_{max}$, for which this uncertainty reaches a maximum value of~3–5\%. Since the geometry and measurement conditions for the reference $^{235}$U$(n,f)$ and the studied $^{236}$U$(n,f)$ reactions were identical during the measurements, the efficiency of registration of fission fragments is the same for the reference and the studied nuclei.

\begin{figure}
\begin{center}
\includegraphics[scale=0.3]{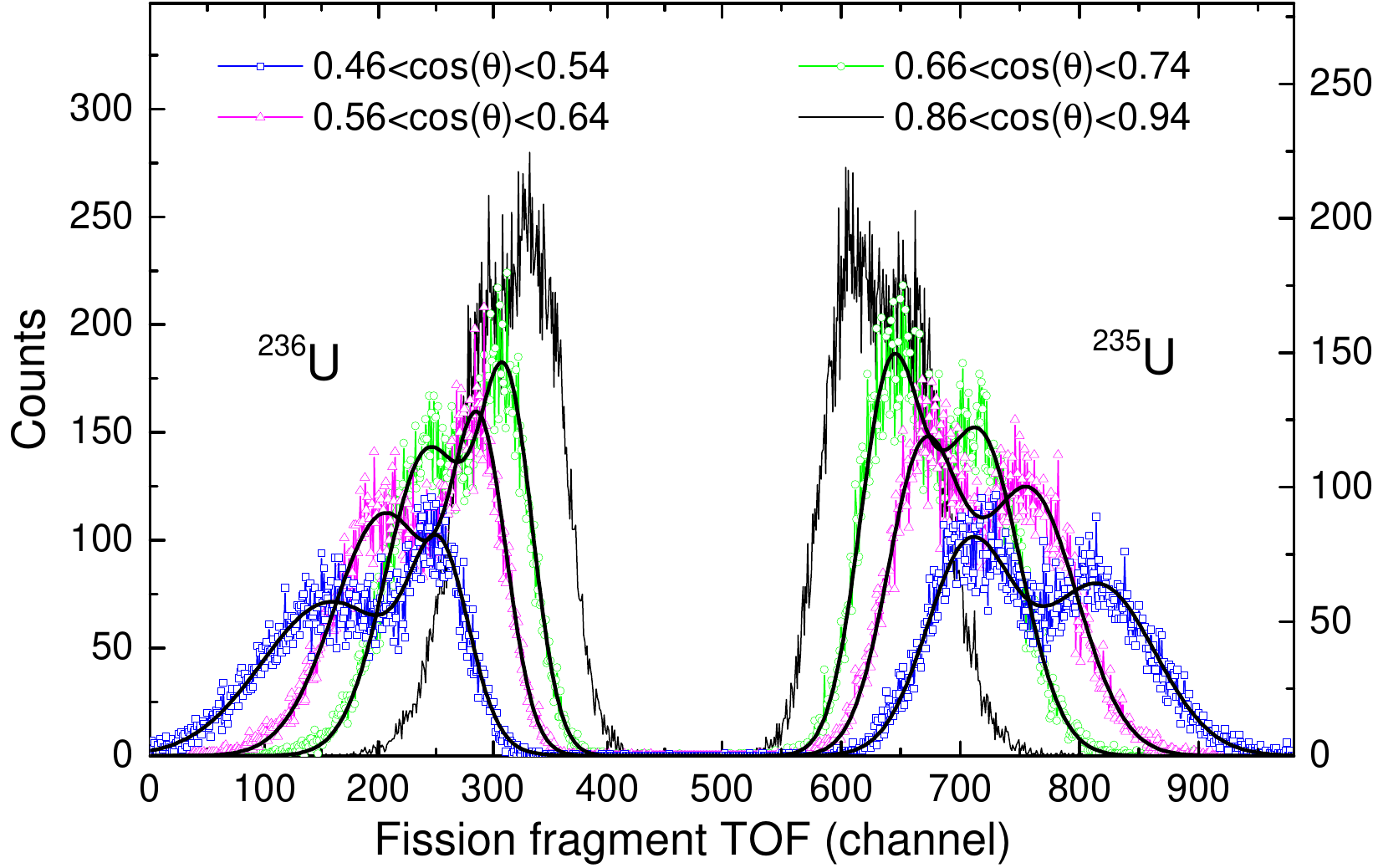}
\caption{Time-of-flight spectrum of the $^{235}$U (to the right of the 480 channel) and $^{236}$U fission fragments (to the left of the 480 channel) depending on the angle between the direction of movement of the fragment and the normal to the plane of the \mbox{MWPCs} electrodes.}
\label{f11}
\end{center}
\end{figure}

During the measurements, the fission fragments of the studied and reference nuclei are recorded simultaneously by the same \mbox{MWPCs}. Therefore, when processing data, it becomes necessary to determine to which nucleus the registered fission fragment corresponds. If we take into account that during the fission of the reference $^{235}$U nuclei, the fragment moves from \mbox{MWPC1} to \mbox{MWPC2}, and when for fission of the studied nuclei  --- from \mbox{MWPC2} to \mbox{MWPC1}, then such identification can be performed by measuring the time of flight of the fragment from the cathode of \mbox{MWPC2} (C2) to the cathode of \mbox{MWPC1} (C1). As an example, Fig.~\ref{f11} shows the time-of-flight spectra of fission fragments obtained for the selected angles of separation of fragments relative to the normal to the plane of the \mbox{MWPCs} electrodes. Two separate groups of events are clearly visible, which correspond to the fission of $^{236}$U$(n,f)$ and $^{235}$U$(n,f)$.

\subsection{\label{s03-2} Energy calibration of the GNEIS spectrometer}

The energy of the neutron causing fission was determined in this work by the time-of-flight method. The signal that serves as the trigger for the start of measurements (digitization of waveforms with the length of 8 $\mu$s) was the signal from the START detector. Using digital processing of waveforms from the cathode of the fragment detector (\mbox{MWPC)} each ``useful'' fission event was assigned a time stamp relative to that trigger, and the time-of-flight spectrum can be formed from such events. The neutron time of flight $t$ was determined by following equation:
\begin{equation}
\label{3}
t=t_{exp}-t_0=t_{exp}-(t_{\gamma}-T_{\gamma})=t_{exp}-t_{\gamma}+\frac{L}{c}\,,
\end{equation}
where $t_{exp}$ --- the time stamp of ``useful'' fission events, $t_0$ --- the time of departure of the neutron from the neutron-producing lead target of the GNEIS spectrometer, $t_{\gamma}$ --- the peak position in the raw time-of-flight spectrum corresponding to fission induced by $\gamma$-quanta (photonuclear fission), and $T_{\gamma}$ --- the time-of-flight of the $\gamma$-quanta from the source target to the target with the isotope under study.

To find out the neutron energy from the time of flight, it is necessary to calibrate the scale ``time of flight -- neutron energy''. Such calibration was carried out using a set of reference points, which consisted of a photofission peak and resonance minima in time-of-flight spectra associated with the position of neutron resonances $E_r$ in the total neutron cross section of lead (the source target material of the GNEIS spectrometer), using the following approximation:
\begin{equation}
\label{4}
t_{exp}=\frac{L}{c}\left(1-\frac{1}{\left(1+E_r/(mc^2)\right)^2}\right)^{-\frac{1}{2}}+t_d.
\end{equation}
Here $L$ is the flight path length, $t_d$ --- the time shift due to the different length of the connecting cables between the detector and the measuring equipment, the response time of the detectors, which give time stamps to measure the time of flight. 

\begin{figure}
\begin{center}
\includegraphics[scale=0.3]{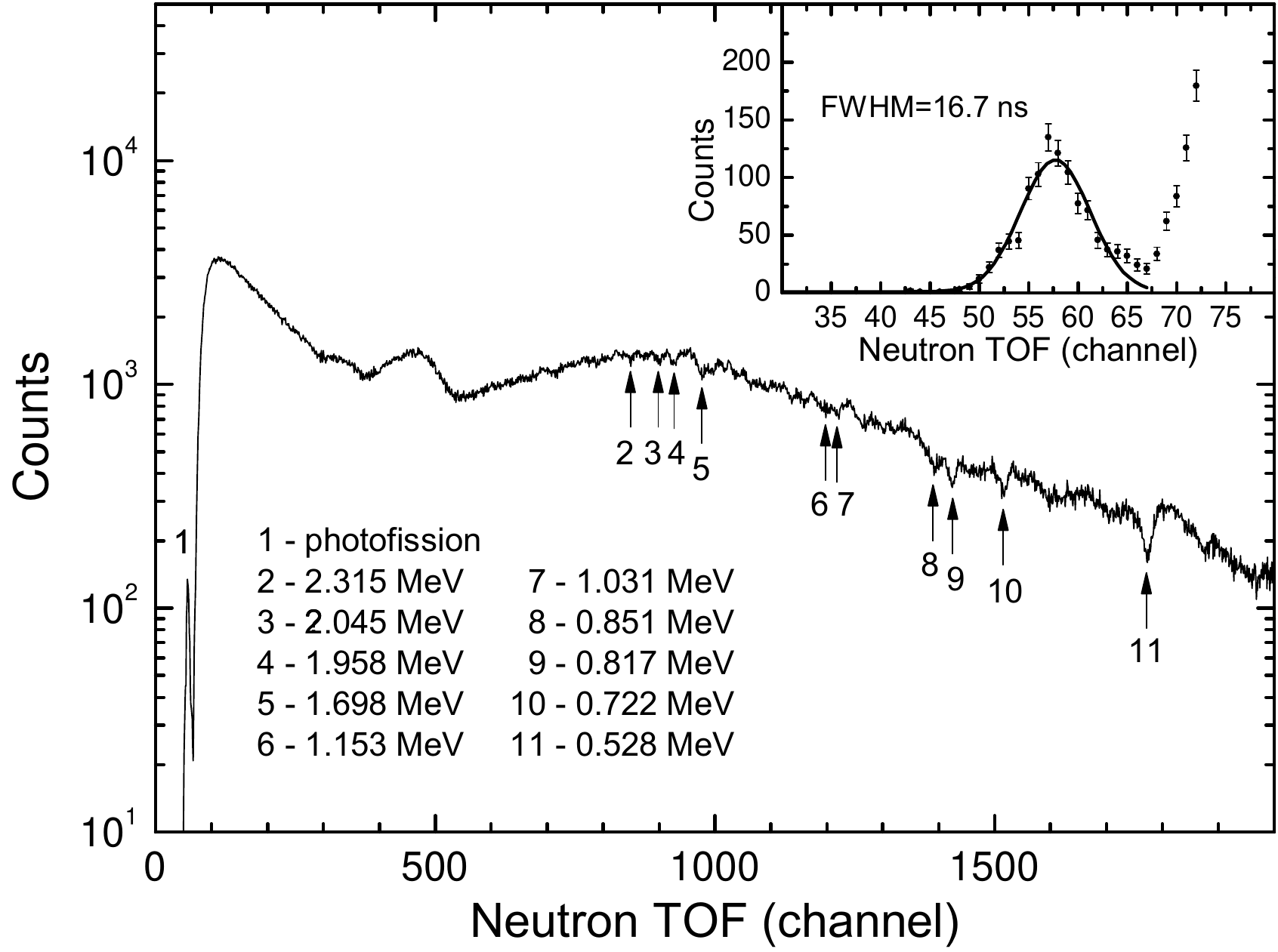}
\caption{The neutron time-of-flight spectrum measured with a $^{235}$U target, the width of the time channel was 2 ns. The inset shows a peak corresponding to photofission.}
\label{f12}
\end{center}
\end{figure}

Fig.~\ref{f12} shows the time-of-flight spectrum obtained using the $^{235}$U target, the arrows indicate the characteristic minima corresponding to the position of neutron resonances in the total neutron cross section of lead and the energy of these resonances. In the region of short flight times, a peak at $N_0\approx 57$ is noticeable, which is associated with the reaction of photonuclear fission of $^{235}$U, the position of which was also used during calibration. The full width of the photonuclear fission peak at half the height is $\approx$ 17 ns. Using Eq.~(\ref{4}), the flight path length $L$ was found to be 36.50(10)~m, and the time shift $t_d$ was 4.8(6.0)~ns. The energy resolution of the spectrometer for the neutron energy range studied in this work is mainly determined by the time uncertainty associated with the proton pulse width, the uncertainty of the linear size of lead target and the time of flight base~\cite{Abrosimov_1985}. In fact, the evaluated energy resolution of the GNEIS spectrometer for neutron energy in the region of 1 MeV is about 1\%, and in the region of 200 MeV --- 12\%.

\subsection{\label{s03-3} Angular distributions of fission fragments}

The angular distribution of fission fragments for the studied nuclei in the center of mass system (c.m.s.), $W_f(\theta,E)$, was determined using the following expression: 
\begin{equation}
\label{5}
\begin{array}{l}
W_f(\theta,E)={}
\\[\bigskipamount]
{\displaystyle \mathstrut\frac{1+\eta(E)}{2}}\,W_{lab,1}(\theta,E)+
{\displaystyle \mathstrut\frac{1+\eta^{-1}(E)}{2}}\,W_{lab,2}(\theta,E),
\end{array}
\end{equation}
\begin{equation}
\label{6}
W_{lab,i}(\theta,E)=\frac{N_i(\theta,E)-N_{SF}}{\varepsilon(\theta)},
\end{equation}
where $W_{lab,1}(\theta,E)$ and $W_{lab,2}(\theta,E)$ are the angular distributions of the fission fragments in the laboratory coordinate system, obtained from the angular distributions $N_1(\cos\theta,E)$ and $N_2(\cos\theta,E)$ measured with the orientation of the installation relative to the direction of motion of neutrons 1 and 2 (Fig.~\ref{f07}), respectively; $N_{SF}$ is the background caused mainly by spontaneous fission in the target substance (note that this component of the background is negligible in these measurements, as it can be seen in the inset of Fig.~\ref{f12}, where there are no events in the region of the time of flight to the left of the peak corresponding to the photonuclear fission of $^{235}$U).
The Eq. (\ref{5}) also includes the ratio:
\begin{equation}
\label{7}
\eta(E)=\frac{\Phi_2 K_2(E)}{\Phi_1 K_1(E)},
\end{equation}
where $\Phi_1$, $K_1(E)$ and $\Phi_2$, $K_2(E)$ are the values of the neutron flux ``at the entrance'' to the fragment detector and the attenuation coefficient of the neutron flux at the location of the target with fissioning nuclei for orientations 1 and 2 of the installation relative to the neutron direction, respectively. The IC monitor with a natural $^{238}$U target located on the neutron beam at a distance of $\approx$ 30 cm from the position-sensitive fragment detector (inset in the upper right corner in Fig.~\ref{f01}) was used as a relative monitor of the neutron beam. In our measurements the $\eta(E)$ was found to be 1.033 with the uncertainty less than 0.3\%.

\begin{figure}
\begin{center}
\includegraphics[scale=0.29]{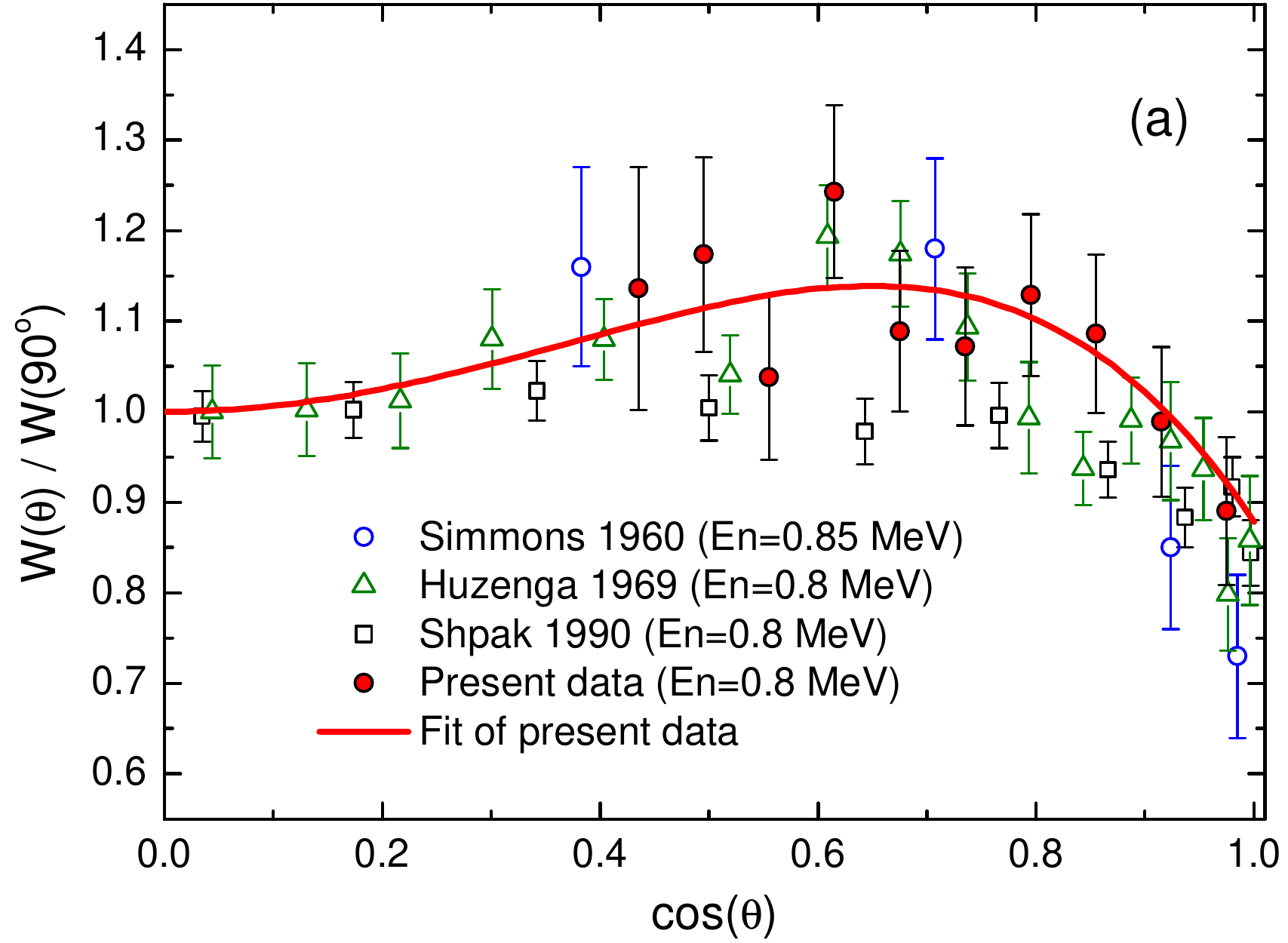}
\includegraphics[scale=0.29]{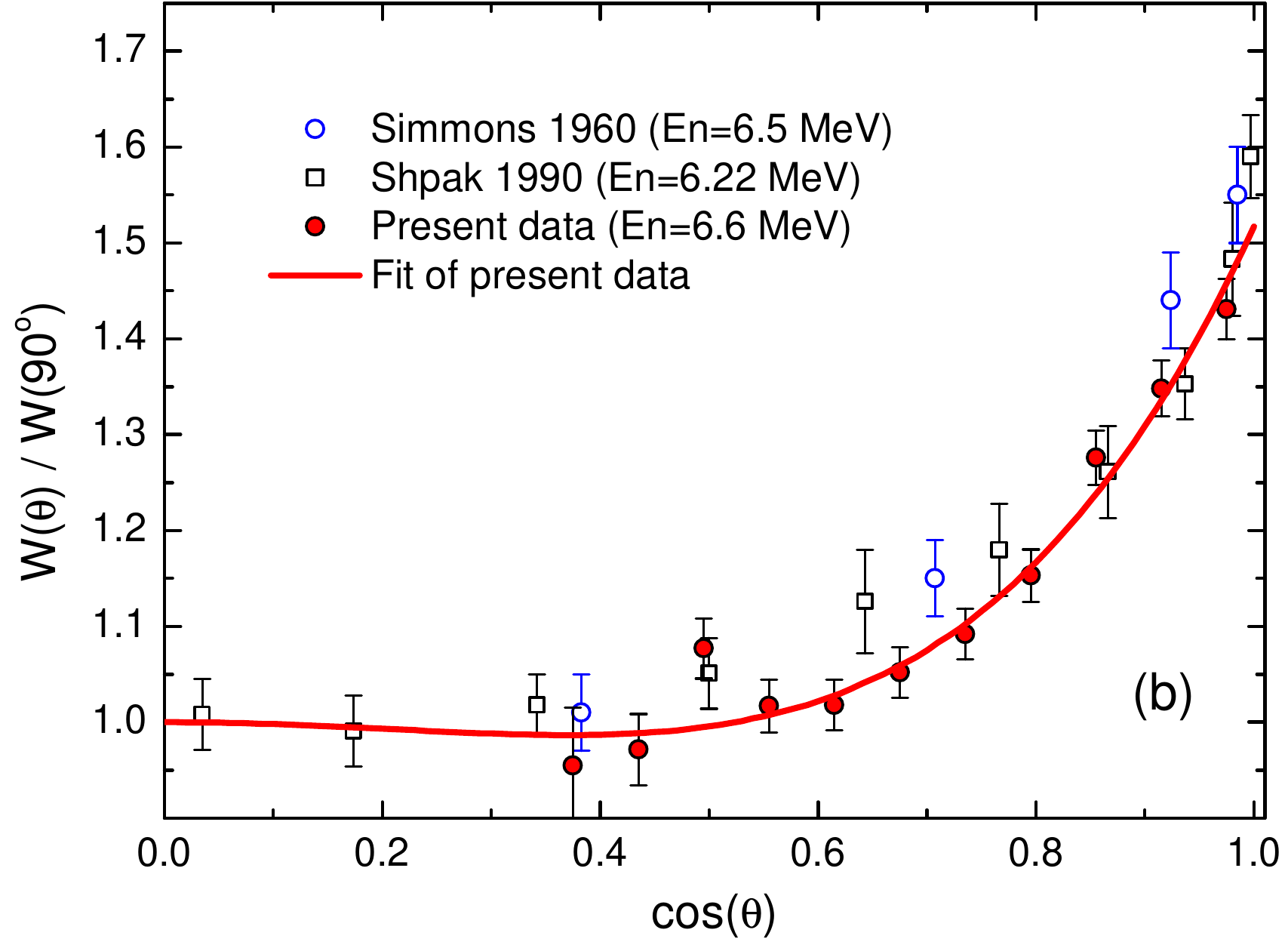}
\caption{Angular distributions of $^{236}$U fission fragments in c.m.s. together with data of other authors \cite{Simmons_1960, Shpak_1991, Huizenga_1969}: (a) --- neutron energy interval from 0.77 MeV to 0.80 MeV; (b) --- neutron energy interval from 6.25 MeV to 6.99 MeV. The error bars in our data represent statistical uncertainties.}
\label{f13}
\end{center}
\end{figure}

For each energy $E$, the angular distributions $W_f(\theta,E)$ measured at $\theta < \theta_{max}$ were approximated by the function $W(\theta,E)$ of the sum of even Legendre polynomials up to the 4th degree, and the step in $\cos\theta$ is 0.01:
\begin{equation}
\label{8}
W(\theta,E)=A_0\left(1+A_2P_2(\cos\theta)+A_4P_4(\cos\theta)\right).
\end{equation}
The result of fitting are the functions $A_0(E)$, $A_2(E)$, $A_4(E)$. The angular distribution $W(\theta,E)$ calculated from the found functions $A_0(E)$, $A_2(E)$, $A_4(E)$ is determined for all angles $\theta$ from $0^{\circ}$ to $90^{\circ}$. For example, the angular distributions of the $^{236}$U fission fragments (in c.m.s.) obtained in this work, or rather the ratio $W(\theta)/W(90^{\circ})$, are shown in Fig.~\ref{f13} together with statistical uncertainties for two selected energy intervals of neutrons causing fission. It should be taken into account that the total uncertainty of the angular distributions also includes the uncertainty of the efficiency of detection of fission fragments, which was discussed above (see Sec. IIA). The data of other authors \cite{Simmons_1960, Shpak_1991, Huizenga_1969} are also presented.

The anisotropy of the angular distribution of fission fragments is determined using the coefficients $A_2$ and $A_4$ for the corresponding Legendre polynomials by the following expression (the $A_0$ is reduced here):
\begin{equation}
\label{9}
\frac{W(0^{\circ})}{W(90^{\circ})}=
\frac{1+A_2+A_4}{1-A_2/2+3A_4/8}.
\end{equation}
In those energy intervals where the coefficient $A_4$ is small in absolute value, the angular distribution of fission fragments is completely determined by the coefficient $A_2$ or the angular anisotropy $W(0^{\circ})/W(90^{\circ})$, which is uniquely expressed in terms of $A_2$.

The uncertainty analysis procedure for anisotropy is similar to the procedure described above for calculating the uncertainty in fission fragment detection efficiency. The parameter varies, and the difference between the original and changed anisotropy values is taken as the uncertainty associated with this parameter. Thus, it was found that the uncertainty of the derived anisotropy due to the geometric uncertainties of the \mbox{MWPCs} and the finite angular resolution is 0.5\%. It should be noted that the accuracy of the anisotropy determination also depends on the reliability of the approximation used for the fitting. In order to estimate the uncertainty related to the fitting procedure the measured angular distributions were fitted by the sum of even Legendre polynomials up to the 6th degree. It turned out that in this case the accuracy of the description does not improve compared to the result obtained using the Eq.~(\ref{f08}) (the parameters $\chi^2$ are close for both approximations). Therefore, the difference between the anisotropy values obtained by fitting with the sum of even Legendre polynomials up to the 4th and 6th degrees was taken as an estimate of the uncertainty related to the fitting procedure. This uncertainty was 1-1.5\%.

The $W(0^{\circ})/W(90^{\circ})$ anisotropy of $^{235}$U fission fragments obtained in the present measurement (the digital data are presented in the Supplemental Material \cite{SM}) is shown in Fig.~\ref{f14} together with statistical uncertainties and compared to the results of other authors \cite{Simmons_1960, Leachman_1965, Nesterov_1967, Ahmad_1979, Musgrove_1981, Androsenko_1982, Meadows_1982, Geppert-Kleinrath_2019, Hensle_2020}. In these experiments were used catcher foil technique \cite{Leachman_1965}, proportional gas counters \cite{Simmons_1960, Nesterov_1967}, semiconductors \cite{Ahmad_1979, Musgrove_1981} and ``track'' detectors \cite{Androsenko_1982}, as well as a gridded ionization fission chamber \cite{Meadows_1982} and a time-projection chamber (TPC) \cite{Geppert-Kleinrath_2019, Hensle_2020}.

In the region of neutron energies below 25~MeV, there is a good agreement of the data within the limits of experimental errors. This fact, in our opinion, may indicate the absence of any significant systematic errors in the obtained anisotropy. Nevertheless, it can be seen that the result obtained in Ref.~\cite{Hensle_2020} lies below the bulk of experimental points available in the international library of experimental nuclear data EXFOR \cite{Otuka_2014}. According to the authors of \cite{Hensle_2020}, this is due to the fact that all previously published data use a normalization at low incident neutron energies in order to determine the detection efficiency of their fission fragment detectors. However, no comparative analysis of the details of various experiments was given. As mentioned above, in our experiment, the fission fragment detection efficiency of an assembly of two \mbox{MWPCs} was calculated by Monte-Carlo method. 

A comparison of the results of other studies performed using neutron sources with a continuous spectrum over a wide energy range up to hundreds of MeV \cite{Vorobyev_2015, Geppert-Kleinrath_2019, Hensle_2020} with the present data is shown in Fig.~\ref{f14}(b). For the neutron energy range above $\approx$~20 MeV, there is a noticeable difference in the present measurements and in our previous work \cite{Vorobyev_2015} (``Vorobyev 2015''), which exceeds their statistical accuracy. We believe that it is due to the fact that in \cite{Vorobyev_2015} ``useful'' fission events were identified only using amplitude spectra of signals from electrodes of \mbox{MWPCs}. Also, an important correction for the ``cross-talk'' effect was not taken into account. Concerning the NIFFTE data \cite{Geppert-Kleinrath_2019, Hensle_2020}, there is an almost constant systematic difference between both of these data sets.

\begin{figure}
\begin{center}
\includegraphics[scale=0.29]{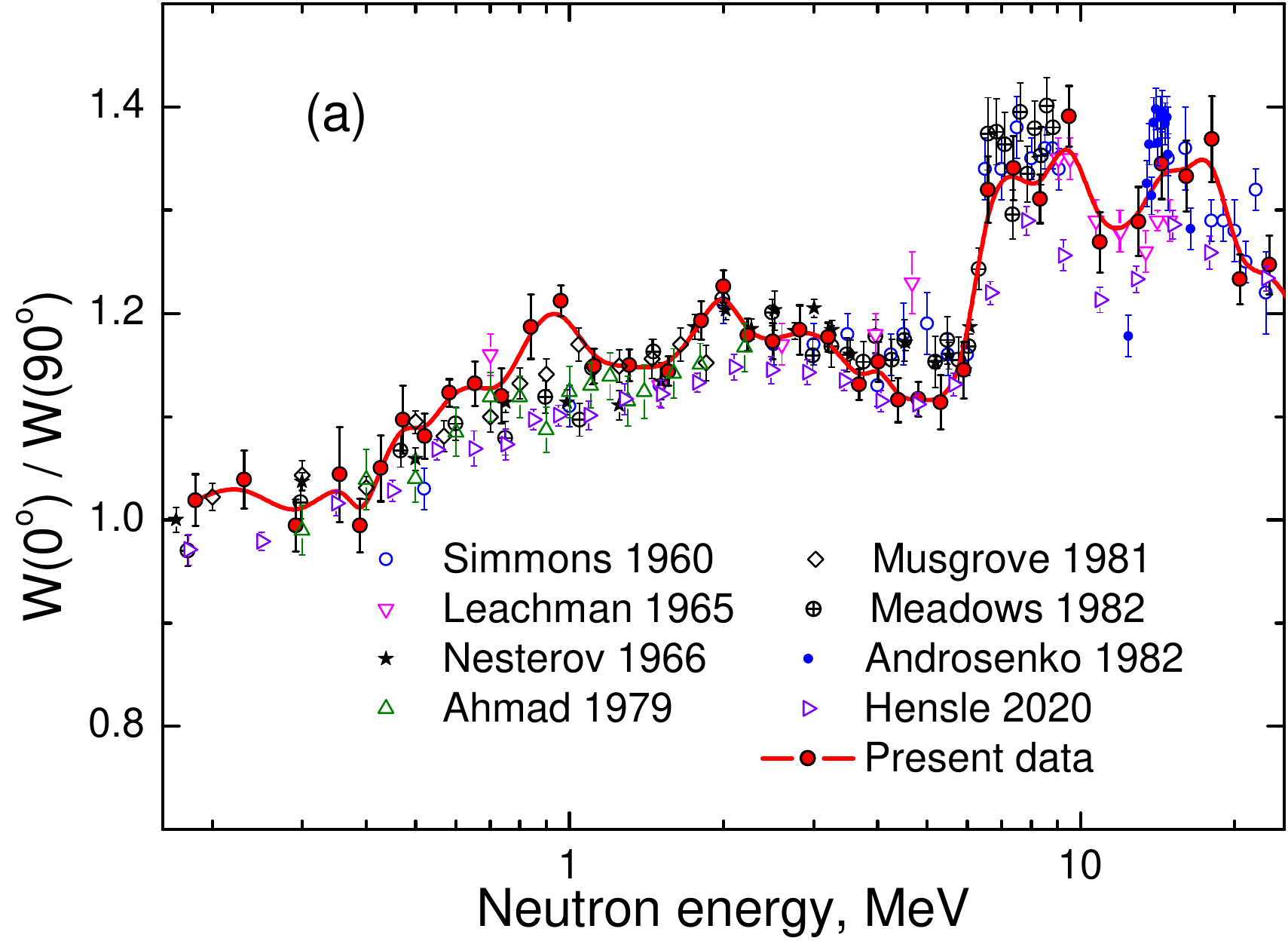}
\includegraphics[scale=0.29]{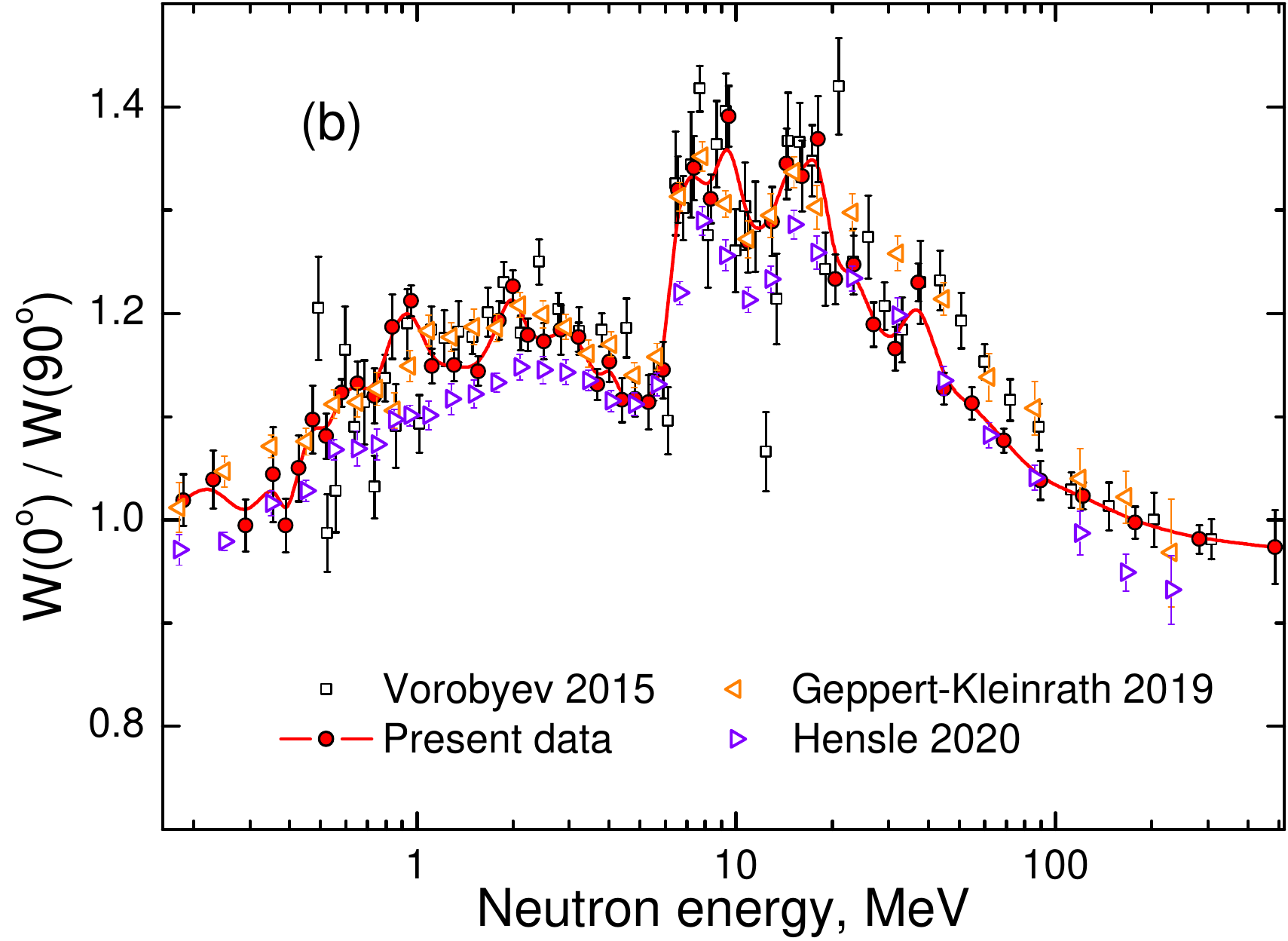}
\caption{The anisotropy of the angular distributions of the $^{235}$U fission fragments, compared with previous results \cite{Simmons_1960, Leachman_1965, Nesterov_1967, Ahmad_1979, Musgrove_1981, Androsenko_1982, Meadows_1982, Geppert-Kleinrath_2019, Hensle_2020}. Panel (a) shows a detailed view for the neutron energy range below 25~MeV. The data obtained using neutron sources with a continuous spectrum over a wide energy range up to hundreds of MeV are shown in panel (b). The error bars in our data represent statistical uncertainties. A solid curve passing through the points is given only for visualization of experimental data.}
\label{f14}
\end{center}
\end{figure}

The anisotropy of the angular distributions of the $^{236}$U$(n,f)$ fission fragments obtained in this work (the digital data are presented in the Supplemental Material \cite{SM}) is shown in Fig.~\ref{f15}.  Comparison of our data with the results of measurements performed by other authors \cite{Simmons_1960, Shpak_1991, Leachman_1965} in the region of neutron energies below 20 MeV demonstrates their good agreement. For neutron energies above 20 MeV, data on the energy dependence of the anisotropy were obtained for the first time. The uncertainty analysis procedure for the $^{236}$U anisotropy was carried out in the same way as for the $^{235}$U case. The uncertainty of the obtained anisotropy due to the geometric uncertainties of \mbox{MWPCs} and the finite angular resolution was 0.5\%. The uncertainty associated with the fitting procedure was 1-2\%.

\begin{figure}
\begin{center}
\includegraphics[scale=0.29]{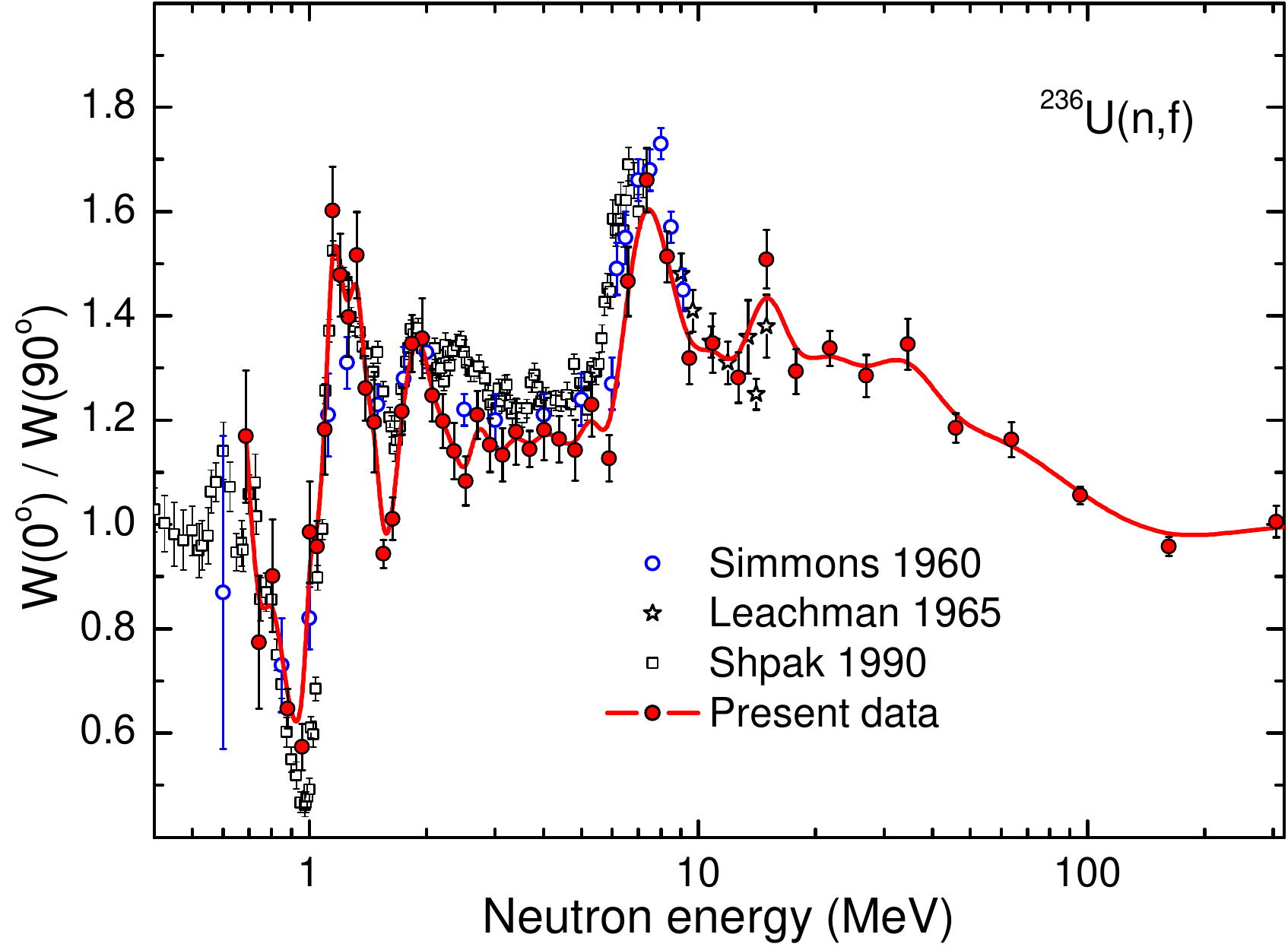}
\caption{The anisotropy of the angular distributions of the $^{236}$U fission fragments obtained in comparison with experimental data of other authors \cite{Simmons_1960, Leachman_1965, Shpak_1991}. The errors are statistical. The solid curve is given only for visualization of experimental data.}
\label{f15}
\end{center}
\end{figure}

\subsection{\label{s03-4} Ratio of the fission cross sections}

\begin{figure}
\begin{center}
\includegraphics[scale=0.3]{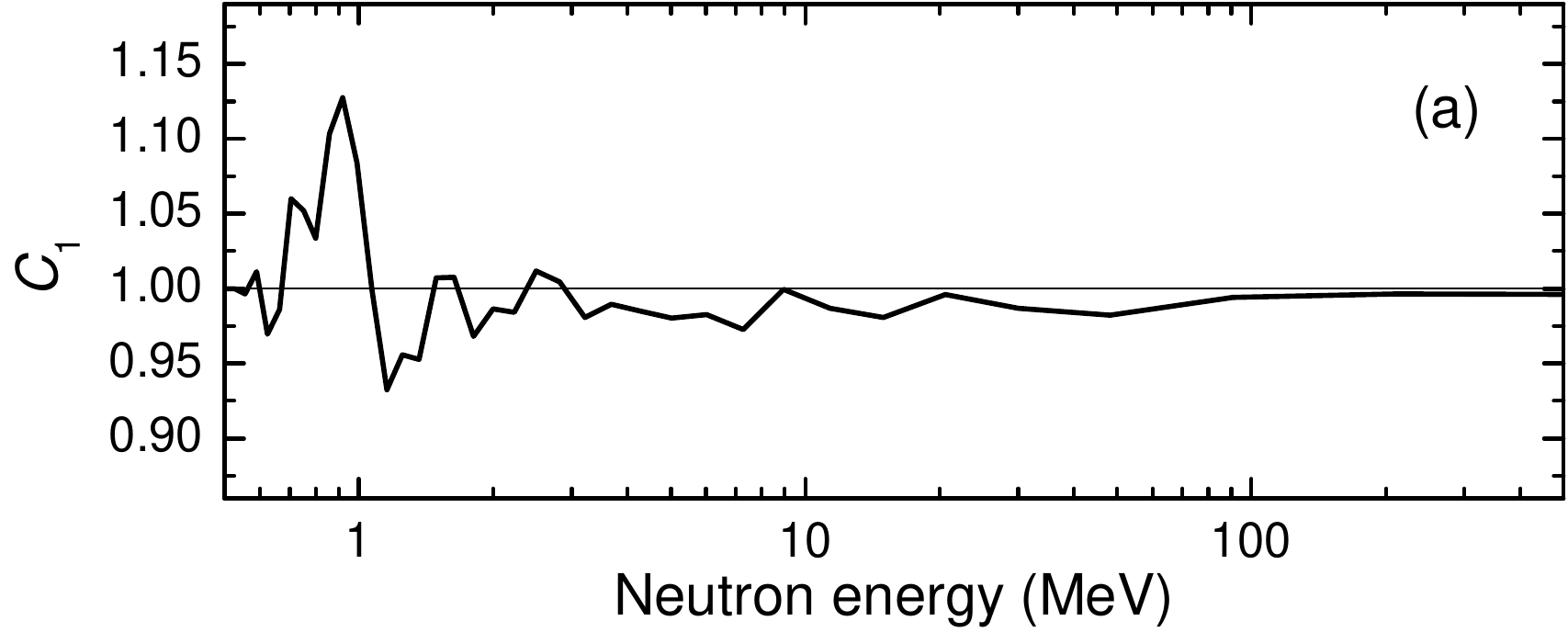}
\includegraphics[scale=0.3]{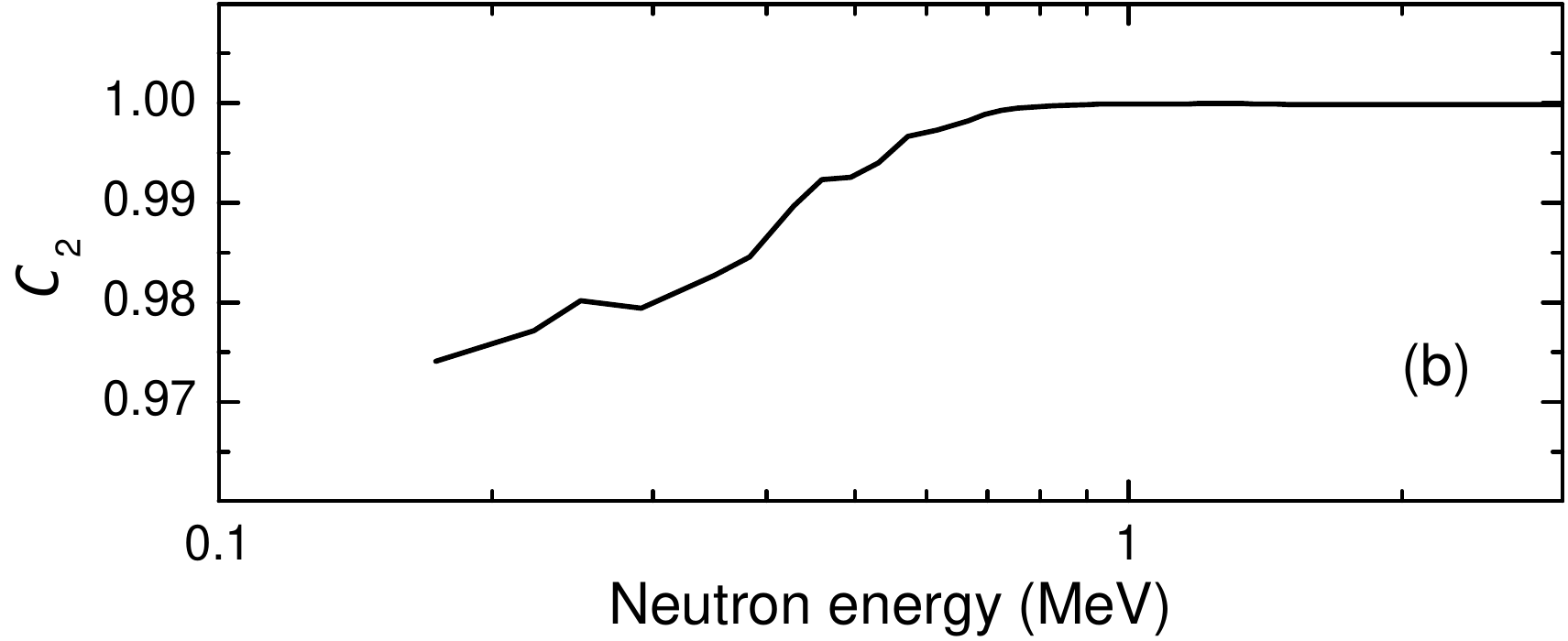}
\caption{Corrections taking into account the anisotropy of the angular distributions of fission fragments $C_1$ --- (a), and the isotopic composition of the targets $C_2$ --- (b).}
\label{f16}
\end{center}
\end{figure}

Taking into account that the targets of the $^{235}$U and $^{236}$U nuclei under study were irradiated with the same neutron flux and had the same geometric dimensions, and the measurements were performed using the same experimental setup, the neutron-induced fission cross section ratio of these nuclei $R(E)=\sigma_f^{U6}/\sigma_f^{U5}$ for incident neutrons with energy $E$ was determined using the following expression:
\begin{equation}
\label{10}
R(E)=\frac{\sum\limits_{\theta < \theta_{\rm max}} W_f^{(U6)}(\theta,E)}
{\sum\limits_{\theta < \theta_{\rm max}} W_f^{(U5)}(\theta,E)}
\frac{N_{U5}}{N_{U6}} C_1(E) C_2(E),
\end{equation}
where $N_{U6}$ and $N_{U5}$ are the number of $^{236}$U and $^{235}$U nuclei in the targets used, respectively; $C_1(E)$ is a correction factor that takes into account the limited solid angle of registration of the position–sensitive fragment detectors and the anisotropy of the angular distributions of the fission fragments (see also Fig.~\ref{f16}):
\begin{equation}
\label{11}
C_1(E)=\frac{\sum\limits_{\theta < \theta_{\rm max}} W^{(U5)}(\theta,E)}
{\sum\limits_{\theta <\theta_{\rm max}} W^{(U6)}(\theta,E)}
\frac{\sum\limits_{\theta < \pi/2} W^{(U6)}(\theta,E)}
{\sum\limits_{\theta < \pi/2} W^{(U5)}(\theta,E)},
\end{equation}
and $C_2(E)$ is a factor that takes into account the isotopic composition of the substance from which the targets $^{236}$U and $^{235}$U were made (see Table~\ref{t2}). This factor is calculated using the following formula:
\begin{equation}
\label{12}
C_2(E)=\frac{P^{(U6)}_{U6}\,\sigma_f^{e(U6)}(E)}
{\sum\limits_k P^{(U6)}_{Uk}\,\sigma_f^{e(Uk)}(E)}
\frac{\sum\limits_k P^{(U5)}_k\,\sigma_f^{e(Uk)}(E)}
{P^{(U5)}_{U5}\,\sigma_f^{e(U5)}(E)},
\end{equation}
where $P^{(U5)}_{Uk}$ and $P^{(U6)}_{Uk}$ are the atomic fraction for the isotope of $Uk$ ($k=4,5,6,8$) in the $^{235}$U and $^{236}$U target substance, respectively, and $\sigma_f^{e(Uk)}(E)$ is the estimated fission cross section for the isotope $Uk$.

As the estimated fission cross sections of the isotopes contained in the targets the following cross sections were taken: for $\sigma_f^{e(U5)}$ --- recommended fission cross section for $^{235}$U \cite{Carlson_2018, Marcinkevicius_2015}, for $\sigma_f^{e(U6)}$ --- the fission cross section of $^{236}$U$(n,f)$ obtained in this work, for $\sigma_f^{e(U8)}$ --- recommended fission cross sections for $^{238}$U \cite{Carlson_2018, Marcinkevicius_2015} for energies above 2 MeV and the fission cross section of $^{238}$U$(n,f)$ from \cite{Behrens_1977} for energies less than 2 MeV, for $\sigma_f^{e(U4)}$ --- the fission cross section of $^{234}$U$(n,f)$ from \cite{Tovesson_2014} for energies less than 200 MeV and $\sigma_f^{e(U5)}$ in the region above 200 MeV. The calculated multiplier $C_2$ is 0.99982-0.99995 in the energy range 1--500 MeV. With a decrease in the neutron energy, the magnitude of this correction increases (see Fig.~\ref{f16}).

It can be seen from the structure of Eq.~(\ref{f10}) that the uncertainty of the obtained ratio, in addition to the statistical accuracy of measurement, also depends on the accuracy of determining the mass of the substance under study (the number of nuclei). As follows from Table~\ref{t1} (right column), the $N_{U6}/N_{U5}$ ratio is 1.658(22).

\section{Results and discussion}

\subsection{\label{s04-1} Present measurements}

The measured ratio of the neutron-induced fission cross section of $^{236}$U and $^{235}$U (the digital data are presented in the Supplemental Material \cite{SM}) is shown in Fig.~\ref{f17} and compared to some measurements performed by other authors \cite{Behrens_1977, Meadows_1978, Goverdovskii_1985, Fursov_1985, Meadows_1988,  Kanda_1986, Lisowski_1992, Sarmento_2011, Tovesson_2014}. Despite the fact that the Lisowski et al. data presented in EXFOR are not final, as the authors themselves noted, and were obtained by digitizing points from a figure published in the conference proceedings \cite{Lisowski_1992}, these data are also presented in the Fig.~\ref{f17} since they are the first data obtained for a wide range of incident neutron energies and the corrections to be made are small. The experimental uncertainties of our ratio are listed in Table~\ref{t4}. The statistical accuracy achieved in this work in the energy range above 0.8 MeV is on average 2.3\% for a given energy bin. The total average systematic error is largely determined by the uncertainty in the thickness of the targets used and amounts to 1.4\%.

\begin{figure}
\begin{center}
\includegraphics[scale=0.3]{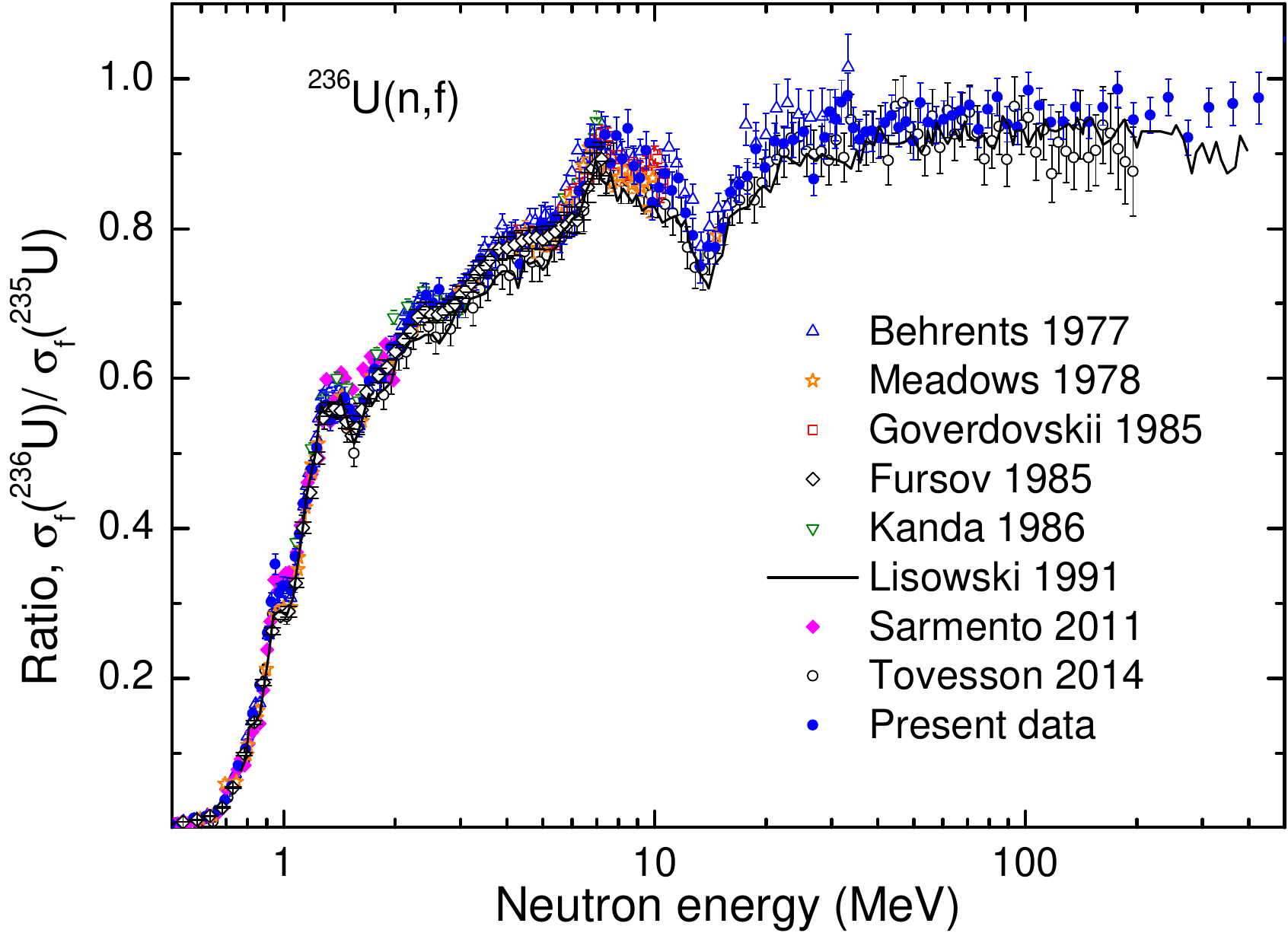}
\caption{Neutron-induced fission cross section of $^{236}$U relative to the one of $^{235}$U. The present data are shown in blue points together with selected EXFOR data \cite{Behrens_1977, Meadows_1978, Goverdovskii_1985, Fursov_1985, Meadows_1988,  Kanda_1986, Lisowski_1992, Sarmento_2011, Tovesson_2014}.}
\label{f17}
\end{center}
\end{figure}

\begin{table}
\caption{\label{t4} Relative experimental uncertainties for fission cross section ratio.}
\begin{center}
\begin{tabular}{|p{4cm}|p{4cm}|}
\hline
\multicolumn{2}{|c|}{Statistical uncertainties} \\
\hline
counting statistic $N_i$ & $25-2.3$\,\% ($0.3-0.8$~MeV) \\
 & $2.3$\,\% (above $0.8$~MeV) \\
\hline
factor $\eta$ & \textless $0.3$\,\% \\
\hline
\multicolumn{2}{|c|}{Systematic uncertainties} \\
\hline
factor $C_1$ & $10$\,\% (\textless $0.4$~MeV) \\
 & $2$\,\% ($0.4-1.0$~MeV) \\
 & $0.5$\,\% (above $1.0$~MeV) \\
\hline
factor $C_2$ & $0.4$\,\% (\textless $0.4$~MeV) \\
 & $0.2$\,\% ($0.4-1.0$~MeV) \\
 & $\approx 10^{-4}$\,\% (above $1$~MeV) \\
\hline
\mbox{MWPCs} geometry & $0.3$\,\% \\
\hline
normalization $N_{U6}/N_{U5}$ & $1.3$\,\% \\
\hline
Total uncertainty & $2.7$\,\%\\
\hline
\multicolumn{2}{|c|}{$^{235}$U standard uncertainty \cite{Marcinkevicius_2015, Carlson_2018}} \\
\hline
 & $1.3-1.5$\,\% (\textless 20 MeV) \\
$\sigma_f^{e(U5)}$ & $1.5-4.8$\,\% ($20-200$ MeV) \\
 & $5-7$\,\% (above 200 MeV)  \\
\hline
\end{tabular}
\end{center}
\end{table}

Figure.~\ref{f18} shows the fission cross section of $^{236}$U (the digital data are presented in the Supplemental Material \cite{SM}) determined from our data for ratio of cross sections (the digital data are presented in the Supplemental Material \cite{SM}). The standard (below 200 MeV) \cite{Carlson_2018} and recommended (300-1000 MeV) \cite{Marcinkevicius_2015} neutron-induced fission cross section of $^{235}$U has been used, that is also shown.
For comparison, the figure also shows the results of earlier works \cite{Behrens_1977, Fursov_1985} and relatively recent measurements \cite{Sarmento_2011, Tovesson_2014}, together with an estimate from the ENDF/B-VIII.0 library \cite{Brown_2018} (the data from \cite{Lisowski_1992} are not shown because there is no information about the measurement errors in this work). All data presented in this figure were determined using the same standard. Within the total error of the data obtained, which also includes the error in determining the number of nuclei in the targets, there is an agreement between the data of this work and the data of other authors in the entire neutron energy range studied.

\begin{figure}
\begin{center}
\includegraphics[scale=0.3]{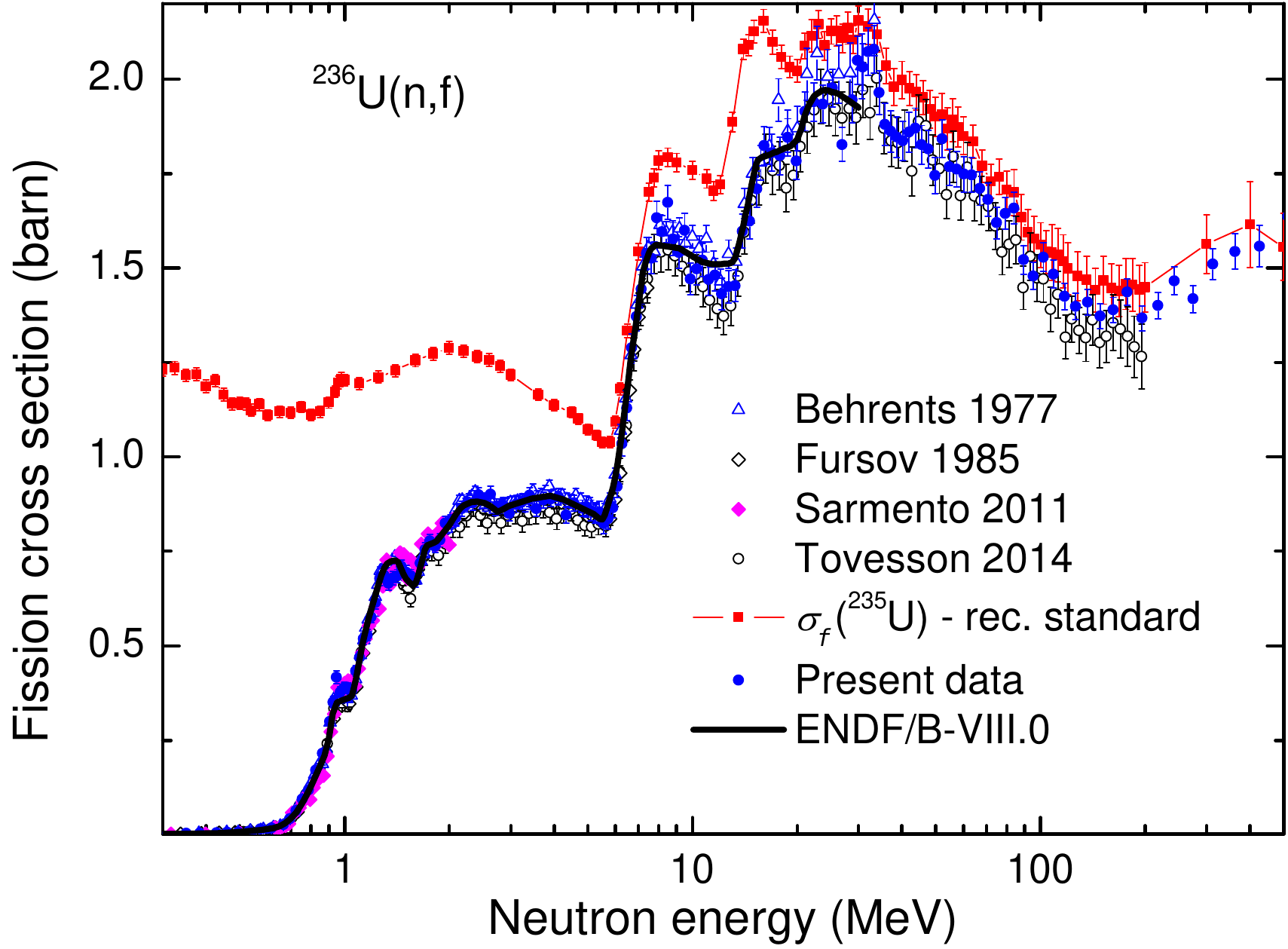}
\includegraphics[scale=0.3]{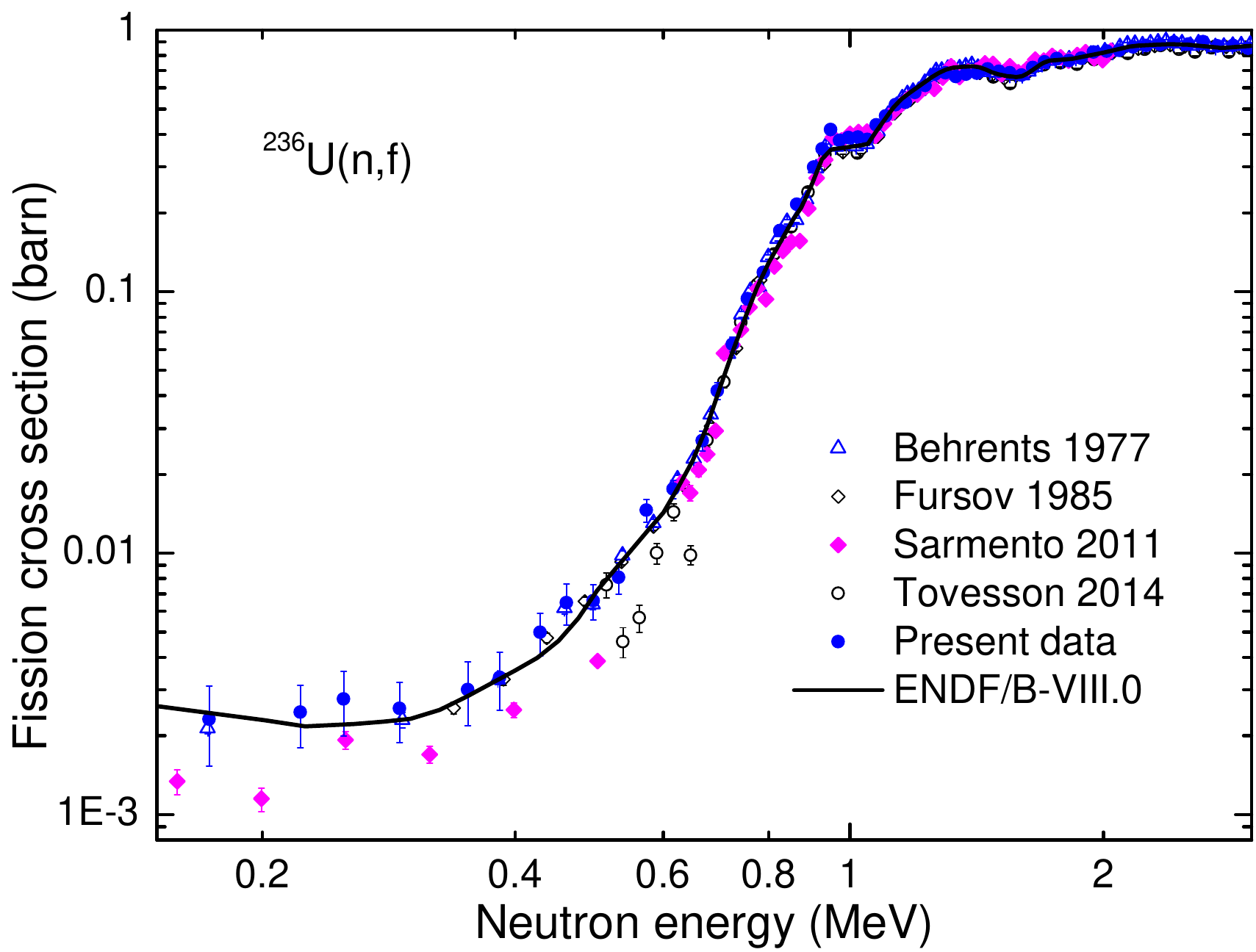}
\caption{Neutron-induced fission cross section of $^{236}$U. The present data are shown in blue points together with selected EXFOR data \cite{Behrens_1977, Fursov_1985, Sarmento_2011, Tovesson_2014}. The standard (below 200 MeV) \cite{Carlson_2018} and recommended (300-1000 MeV) \cite{Marcinkevicius_2015} neutron-induced fission cross section of $^{235}$U are shown in red squares. The error bars plotted here represent a total uncertainty of the measurements (the error of the $^{235}$U standard is not included).}
\label{f18}
\end{center}
\end{figure}

\begin{figure}
\begin{center}
\includegraphics[scale=0.3]{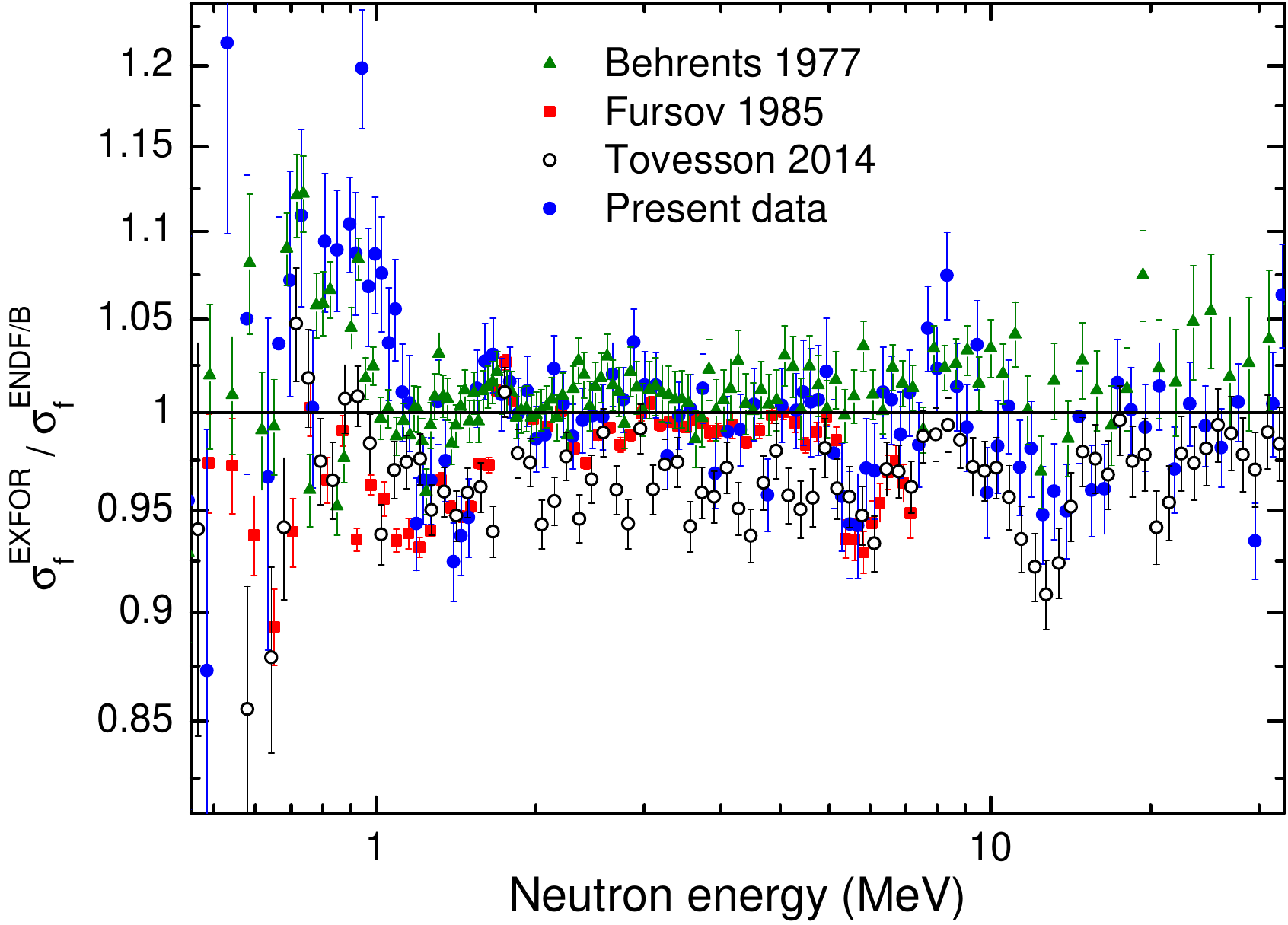}
\caption{Deviation of the discussed data sets shown in Fig.~\ref{f18} with respect to the ENDF/B-VIII.0 evaluation. The error bars plotted here only represent the statistical uncertainty of the measurements.}
\label{f19}
\end{center}
\end{figure}

Although the data obtained in different studies are in good agreement, there are differences between them. They are most clearly shown in Fig.~\ref{f19}, where the deviations of the experimental data previously given in Fig.~\ref{f18} with respect to the ENDF/B-VIII.0 evaluation are presented. One can see that in the neutron energy range from $\approx$ 1.0 to 30 MeV, the neutron-induced fission cross section of $^{236}$U from the ENDF/B-VIII.0 library generally correctly describes the available experimental data, since the ratio of experimental data and the ENDF/B-VIII.0 evaluation is on average constant and close to 1. The largest deviation from 1, which exceeds the statistical accuracy of the data, is observed in the region of a sharp change in the fission cross section corresponding to the opening of the 1st, 2nd and 3rd chances. We also note that, on average, there is a difference in absolute value between all experimental data, which is largely related to the accuracy of normalization to the number of nuclei and corrections for the efficiency of detecting fission fragments.

In the present work, the normalization accuracy is 1.4\%, while in \cite{Behrens_1977, Fursov_1985} and \cite{Sarmento_2011, Tovesson_2014} it is ~1.0\% and ~3.0\%, respectively. The largest average shift relative to unity is noticeable for the data of \cite{Tovesson_2014}, which practically coincide with the data of \cite{Lisowski_1992} and are lower than other experimental data and the ENDF/B-VIII.0 evaluation by ~4\% on average. In a recent work \cite{Ren_2020} (the numerical data obtained in this work are not available in the EXFOR library), it is also noted that the shape of the curve of the measured fission cross section of $^{236}$U$(n,f)$ agrees with the estimate from ENDF/B-VIII.0, and the average bias relative to the ENDF/B-VIII.0 evaluation is minus 2.8\% (with an uncertainty of the normalization coefficient of 1.5\%).

\begin{figure}
\begin{center}
\includegraphics[scale=0.3]{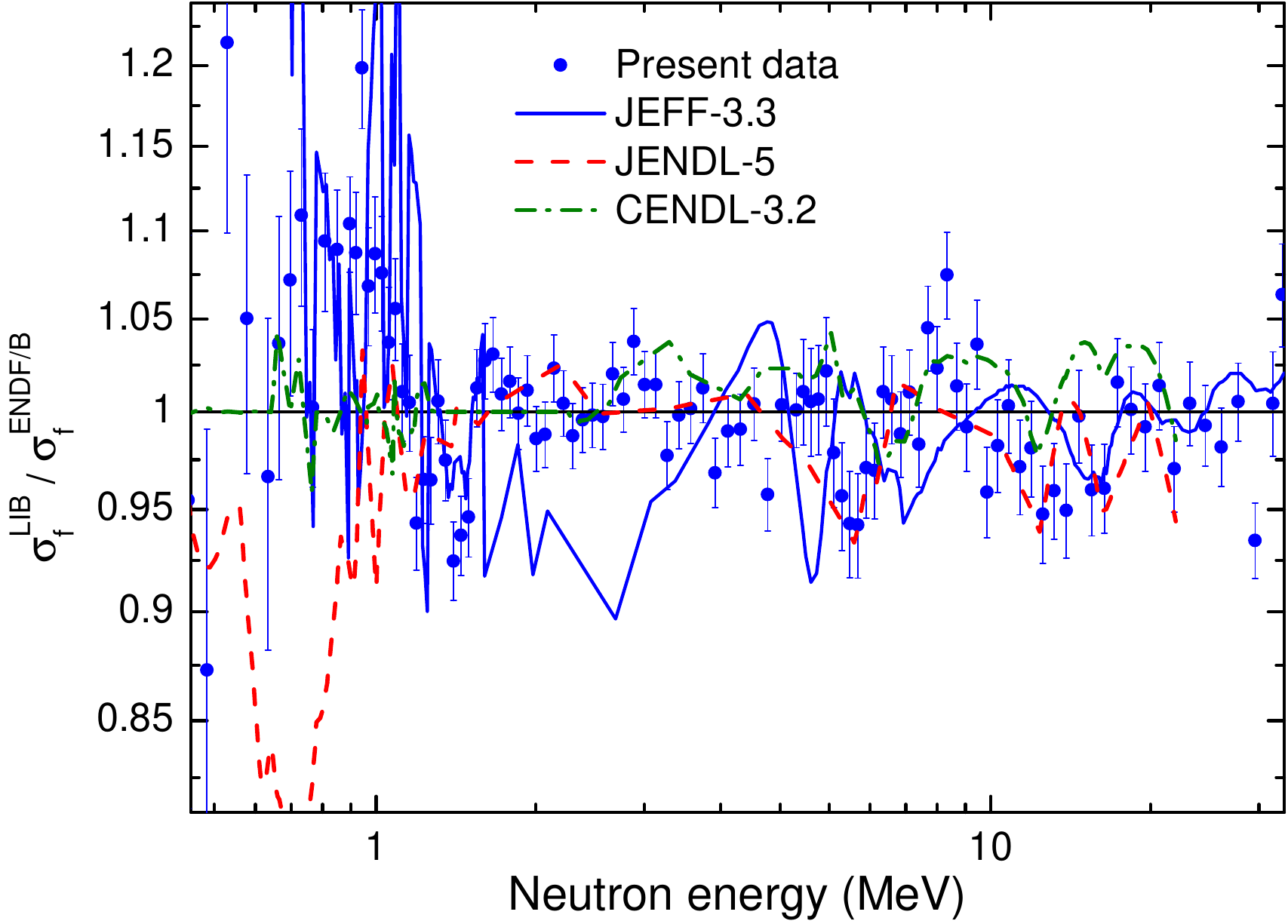}
\caption{Deviation of the fission cross section of $^{236}$U$(n,f)$ evaluated by different national libraries with respect to the ENDF/B-VIII.0 evaluation.}
\label{f20}
\end{center}
\end{figure}

Fig.~\ref{f20} shows the variation that exists between different national libraries of evaluated data: JEFF-3.3 \cite{OECD_2018}, JENDL-5 \cite{Iwamoto_2020}, CENDL-3.2 \cite{Ge_2020}, and ENDF/B-VIII.0 (note that the ENDF/B-VIII.0 evaluation in the neutron energy range below 20 MeV coincides with the ROSFOND-2010 evaluation \cite{Zabrodskaya_2007}). In the energy range of 1–20 MeV, all evaluations agree with each other within $\approx$ 3\%, with the exception of the European evaluation JEFF-3.3, according to which the fission cross section of $^{236}$U in the neutron energy range from 1.5 MeV to 3 MeV is underestimated by 5-8\% relative to other estimates. For neutron energies less than 1.0 MeV it is seen that the JENDL-5 evaluation on average tend to be about 10\% lower than other ones.  

\subsection{\label{s04-2} Discussion of the cross section $^{236}$U$(n,f)$}

To date, quite advanced methods have been developed for calculating various reactions caused by the interaction of nuclei with light particles, including neutrons, in a wide range of collision energies --- see, for example, the review \cite{Capote_2009} about the Reference Input Parameter Library (the RIPL-3 version). Such calculations, in particular, can be performed using the open software package TALYS \cite{Koning_2008}. Because the parameters (used by default) of the nuclear models included in TALYS are adjusted from the results of many tests, this software package provides a realistic description of a wide range of nuclear reactions
for energies up to 200 MeV. But some caution is required when describing nuclear fission when additional parameters are used to model both fission barriers and transition states at the fission barriers. The cross sections of nuclear fission by light particles, in particular neutrons, especially at relatively high collision energies, are described by the TALYS program (as well as other similar programs), as a rule, not very reliably.

However, the larger the array of experimental data on nuclear fission cross sections, the greater the possibilities for testing fission models and selecting the correct values for  their parameters. In this case, the fission cross sections at high collision energies are of particular interest, since they are formed as a result of the addition of contributions from fissions of the 1st, 2nd, 3rd, etc. chances, i.e. turn out to be sensitive to the parameters of barriers and transition states of several fissionable nuclei at once.

\begin{figure}
\begin{center}
\includegraphics[scale=0.3]{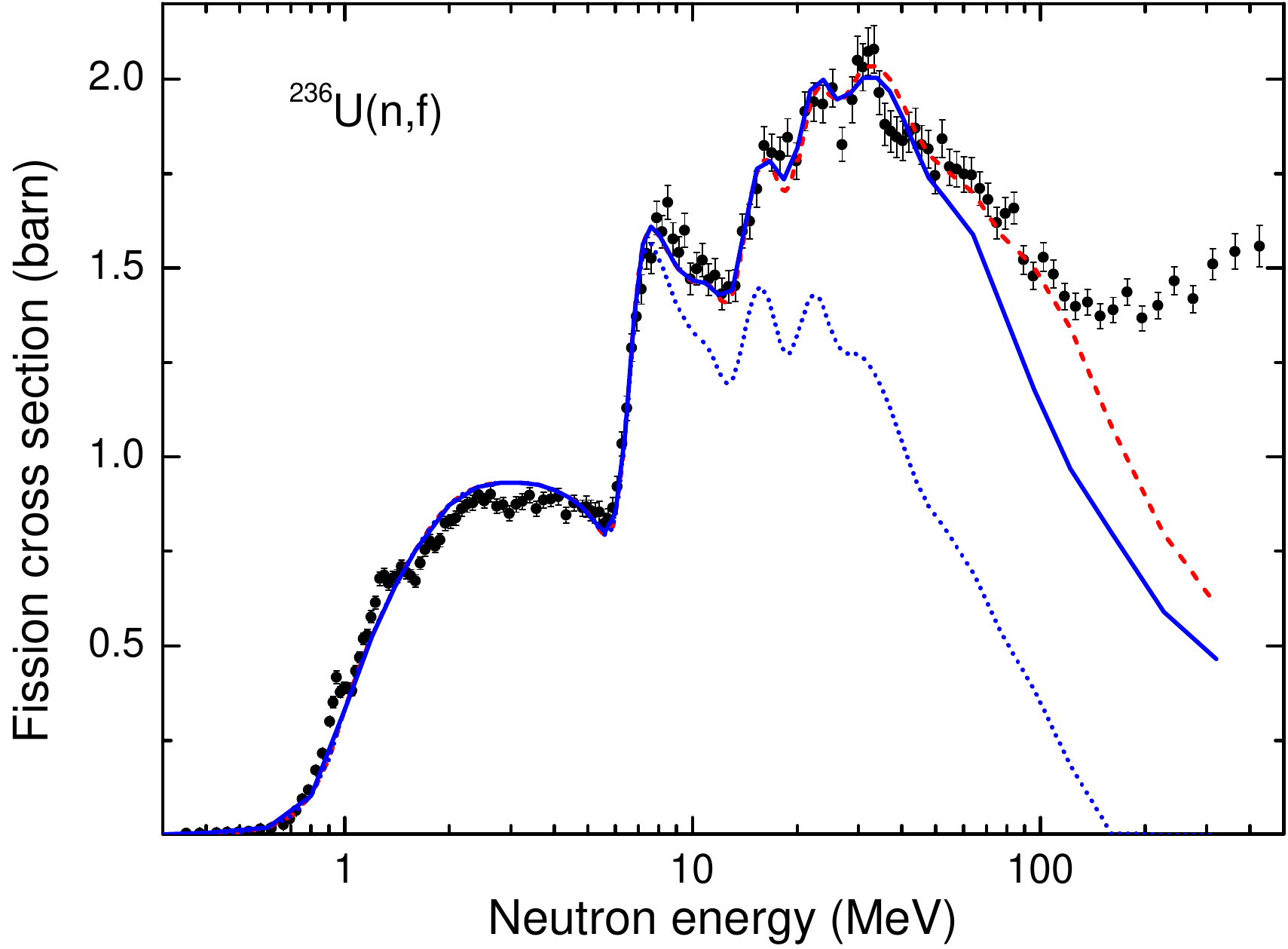}
\caption{Neutron-induced fission cross section of $^{236}$U. The results of present measurements are shown by points together with calculations (solid and dashed curves - see explanations in the text of the article). The dotted curve is the calculated contribution to the fission cross section of the primary compound nucleus and nuclei formed at all stages of its decay.}
\label{f21}
\end{center}
\end{figure}

With this in mind, we attempted to describe the measured cross section of $^{236}$U nuclear fission by neutrons in the range from 0.5 to 300 MeV using standard tools provided by the TALYS-1.9 program. Since at energies of incident neutrons up to 40-50 MeV the main contributions to the cross section are made by fission from the 1st to the 6th chances, a reasonable description in this region can be achieved by changing (relative to the default values) some parameters that determine the fission barriers and the density of transition states above these barriers for nuclei from $^{237}$U to $^{232}$U (the parameters of all other nuclei at this stage of the analysis were assumed to be equal to the default values) --- see Table~\ref{t5} . The resulting cross section is shown in Fig.~\ref{f21} by a solid line.

\begin{table*}
\caption{\label{t5} Heights $B$~(MeV), width parameters $\hbar\omega$~(MeV), factors $R_{tm}$ and $K_{rc}$ for the 1st and 2nd fission barriers of nuclides $^{237-232}$U depending on the mass number $A$, used to describe the fission cross section and the angular anisotropy of fragments in the ${}^{236}$U$(n,f)$ reaction. If the parameter values differ from the default value, then the latter is indicated next to the former in brackets.}
\begin{center}
\begin{tabular}{|c|c|c|c|c|c|c|c|c|}
\hline
 & \multicolumn{4}{c|}{barrier 1} & \multicolumn{4}{c|}{barrier 2} \\
\hline
$A$ & $ B$ & $ \hbar\omega$ & $ R_{tm}$ & $ K_{rc}$ & $ B$ & $ \hbar\omega$ & $ R_{tm}$ & $ K_{rc}$ \\
\hline
237  & $6.3\, (6.4)$  & $0.9\, (0.7)$ & $0.7\, (0.6)$ & $1.5\, (1.0)$ & $6.15$ & $0.5$ & $1.0$ & $8.0\, (1.0)$ \\
\hline
236  & $5.0$  & $0.9$ & $0.6$ & $4.0\, (1.0)$ & $5.95\, (5.67)$ & $0.5\, (0.6)$ & $3.0\, (1.0)$ & $7.0\, (1.0)$ \\
\hline
235 & $5.25$ & $0.7$ & $0.6$ & $1.0$ & $5.7\, (6.0)$ & $0.5$ & $1.0$ & $2.0\, (1.0)$ \\
\hline
234 & $4.8$ & $0.9$ & $0.6$ & $1.0$ & $5.4\, (5.5)$ & $0.6$ & $1.0$ & $4.0\, (1.0)$ \\
\hline
233 & $4.35$ & $0.8$ & $0.6$ & $1.0$ & $5.2\, (5.55)$ & $0.5$ & $4.0\, (1.0)$ & $1.0$ \\
\hline
232 & $4.9$ & $0.9$ & $0.6$ & $1.0$ & $5.7\, (5.4)$ & $0.6$ & $1.0$ & $1.0$ \\
\hline
\end{tabular}
\end{center}
\end{table*}

In addition to the heights and widths of the barriers, the Table~\ref{t5} lists the parameters $R_{tm}$ and $K_{rc}$, which significantly affect the density of transition states. The values of these parameters are determined by the TALYS keywords ``Rtransmom'' and ``Krotconstant''.

To clarify the meaning of $R_{tm}$ and $K_{rc}$, we note that TALYS calculates by default the density of transition states (or levels) $\rho^{tot}(E_x)$ at each fission barrier in the Gilbert--Cameron model, i.e. $\rho^{tot}(E_x)$ is given by the function $\exp [(E_x-E_0)/T]$ for $E_x<E_M$ and is assumed to be equal to the Fermi-gas level density for $E_x>E_M$, while $E_0$, $T$, and $E_M$ determined, firstly, by the continuity and smoothness conditions for the function $\rho^{tot}(E_x)$ and, secondly, by the number of discrete transition states $N_U-N_L$ with excitation energies $E_L<E_x\le E_U<E_M$. Thus, the larger the number $N_U-N_L$ for fixed $E_L$ and $E_U$, the higher the level density for $E_x>E_M$. Now note that the discrete states belong to the rotational bands, and the distances between the states are inversely proportional to the transverse moment of inertia $\tilde {\cal J}_{\perp}$ of the nucleus at the barrier. It is determined by the product of $R_{tm}$ and the parameter
\begin{equation}
\label{14}
{\cal J}_{\perp}={\cal J}_0\left(1+\frac{\beta_2}{3}\right),
\end{equation}
where ${\cal J}_0=2m_nR^2A/5$ is the moment of inertia of a spherical nucleus with the number of nucleons $A$ and radius $R=1.2 A^{1/3}$~fm, $m_n$ is the neutron mass, and $\beta_2$ is the nuclear deformation parameter (in TALYS by default $\beta_2$ is equal to $0.6$ for the 1st barrier and $0.8$ --- for the 2nd one). Thus, by increasing $R_{tm}$, we increase $N_U-N_L$, thereby increasing $\rho^{tot}(E_x)$ for all values of $E_x$. As for $K_{rc}$, this is an additional ``adjustment'' factor for the collective enhancement of the density of transition states at $E>E_M$; by default (as it can be seen, in particular, from the Table~\ref{t5}), it is always equal to one. Thus, by increasing $K_{rc}$, we increase $\rho^{tot}(E_x)$ at $E>E_M$.  It should be noted that the fission cross section reacts significantly differently to changes in the parameters $R_{tm}$ and $K_{rc}$.

It is clear that $R_{tm}$ and $K_{rc}$ are far from the only parameters that can be used to change the density of transition states. The level density in TALYS is set using a variety of parameters, perhaps the most important of which is the level density parameter $a$ (see \cite{Koning_2017}). However, we use default values for all these parameters, assuming, in particular, that the Fermi-gas density of levels is a very good approximation, tested on a huge variety of isotopes, both spherical and deformed (albeit with relatively small deformations). At the same time, the parameters $R_{tm}$ and $K_{rc}$ are specific for highly deformed nuclei, and their values cannot be established in any other way than by studying the features of nuclear fission.

At the same time, however, we assumed that the deviations of the parameters from the default values would be small. Some of the expectations were justified: for example, corrections, as a rule (the only exceptions are the isotopes $^{237}$U and $^{236}$U), refer to parameters associated with a higher barrier for a given nucleus. In general, however, there are more surprises.

Indeed, as can be seen from the Table~\ref{t5}, the deviations from the default parameters are relatively small only for the $^{235}$U and $^{232}$U isotopes. Take, however, the $^{237}$U isotope. If the parameters on the first (higher) barrier differ only slightly from the default values, then on the second barrier the value of $K_{rc}$ has to be increased from $1.0$ to $8.0$. In our opinion, this indicates some significant error in the model. For example, this one: in accordance with the default parameters (taken from \cite{Capote_2009}), the $^{237}$U nucleus at the first fission barrier is considered to be axially asymmetric (with an increased collective enhancement of the level density), while at the second barrier --- mass-asymmetric, but axially symmetric (with a reduced collective enhancement). The increased coefficient $K_{rc}$ on the second barrier $^{237}$U possibly compensates for this inaccuracy.

We treat other results obtained in a similar way. A significant deviation of the parameter from the default value indicates rather a defect inherent in the model than the real value of this parameter. Thus, in particular, there are no physical grounds to expect that the moments of inertia of nuclei at the barriers are 3-4 times greater than the rigid-body values (the $R_{tm}$ parameter was introduced, rather, to describe a possible decrease in the moments of inertia relative to the rigid-body values due to the effect nuclear superfluidity). Thus, the increased values of $R_{tm}$ (see Table~\ref{t5}) seem to signal a lack of understanding of the structure of the discrete spectra of transition states at the barriers. It seems to us that the value of our analysis lies precisely in identifying the above inconsistencies. Overcoming them requires separate efforts.

Note now that above 50 MeV the solid curve in Fig.~\ref{f21} is systematically below the measured values of the fission cross section. In this region, it is not enough to take into account the following (7th, 8th, etc.) chances of fission of U isotopes. Here, the fission of residual nuclei, which are formed during the emission of not only neutrons, but also charged light particles, primarily protons, becomes significant. These nuclei, as a rule, are far from the stability line and have short lifetimes. Accordingly, there is no experimental information on barriers for these nuclei, which can be obtained from the $(n,f)$ reaction at relatively low neutron energies. All such barrier information obtained from $(n,f)$ for relatively stable actinide isotopes is presented in RIPL-3 and, in particular, is given in Table~XI of the review \cite{Capote_2009}. It can be said that in the $^{236}$U$(n,f)$ reaction at energies above 50 MeV, the fission cross section is essentially determined by nuclei, for which there is no experimental information on fission barriers in RIPL-3. Accordingly, in TALYS, by default, the fission barriers of such nuclei are assigned either the values calculated in \cite{Mamdouh_2001}, or, in the absence of such values, the barriers are calculated according to the method proposed in \cite{Cohen_1974}.

Since such calculations can systematically overestimate barriers, we have reduced all barriers for which there is no experimental information in RIPL-3 by 20\%. The cross section for the fission of $^{236}$U nuclei by neutrons calculated under this additional assumption is shown in Fig.~\ref{f21} by a dashed line. It can be seen that the result agrees with the measured cross section up to an energy approximately equal to 120 MeV. Thus, it is possible that the calculated fission barriers are indeed overestimated, but, of course, not literally by 20\% each, but, conditionally, by an average of 20\%. Above 120 MeV, there is again a discrepancy between the calculated and measured cross sections. To understand the reasons for this, obviously, separate efforts are required.

\subsection{\label{s04-3} Angular anisotropy of fragments in the reaction $^{236}$U$(n,f)$}

In accordance with the A.~Bohr's transition state model \cite{Bohr_1956, Bohr_1998}, the angular distribution of fission fragments of nuclei with aligned spins is determined by the probability distribution of fission by the projection $K$  of the spin of the fissioning nucleus onto the deformation axis. In particular, the uniform distribution over $K$ leads to isotropy of fragment emission with respect to the spin alignment axis. On the contrary, fission through some unique value of $K$ leads to a characteristic anisotropic angular distribution.

Since a fissioning nucleus with a relatively low excitation energy passes over the barrier, being in one of the discrete transition states with a certain projection $K$ of spin on the deformation axis, then for such a nucleus the angular anisotropy of the fragments should be especially noticeable. This is confirmed by numerous experiments --- see, in particular, Fig.~\ref{f15}, where the angular anisotropy $W(0^{\circ})/W(90^{\circ})$ changes significantly at energies from 0.5 to 3 MeV. However, analysis of such experiments has not yet yielded any significant information about the positions and characteristics of transition states of nuclei above barriers; all the data on these states included in \mbox{RIPL-3} \cite{Capote_2009} are extracted from the analysis of the energy dependence of the total fission cross sections.

But the case where the fissioning nucleus has a relatively high excitation energy over a higher barrier has been studied much better. Here one can use a statistical approach (see, for example, the monograph \cite{Vandenbosch_1973}), according to which the probability of fission through a state with projection $K$ is proportional to the density $\rho(J,\pi,K)$ of transition states with projection $K$. For given $J$ and $\pi$, this density can be represented as
\begin{equation}
\label{15}
\rho(K)\sim e^{-K^2/2K^2_0},\quad K^2_0=\frac{T{\cal J}_{\rm eff}}{\hbar^2}=
\frac{T{\cal J}_{\perp}{\cal J}_{\|}}{\hbar^2({\cal J}_{\perp}-{\cal J}_{\|})}\,,
\end{equation}
where ${\cal J}_{\perp}$ (\ref{14}) and
\begin{equation}
\label{17}
{\cal J}_{\|}={\cal J}_0\left(1-\frac{2\beta_2}{3}\right)
\end{equation}
are the transverse and longitudinal moments of inertia of the nucleus at a higher fission barrier, $T$ is the temperature of the nucleus. If the moments of inertia, like the temperature $T$, smoothly depend on the excitation energy, then the angular anisotropy of the fragments must also be a smooth function of the energy. Therefore, based on Fig.~\ref{f15}, we can assume that in the $n+^{236}$U reaction, the applicability region of the statistical model for the formation of angular anisotropy of fragments begins with an energy close to 3 MeV.

Previously, calculations of the angular anisotropy of fragments in multi-chance fission were performed by the authors of \cite{Ryzhov_2005} and ourselves. In \cite{Ryzhov_2005} the reactions $^{232}$Th$(n,f)$ and $^{238}$U$(n,f)$ are considered for incident neutron energies up to 100 MeV. Our analysis was devoted to the reactions $^{237}$Np$(n,f)$ \cite{Vorobyev_2019, Barabanov_2020, Barabanov_2021} and $^{240}$Pu$(n, f)$  \cite{Vorobyev_2020} for neutrons with energies up to 200-300 MeV. Such an analysis cannot be performed using programs such as TALYS, since they lack the ability to calculate the angular distribution of fragments. Nevertheless, we managed to use the TALYS-1.9 program having improved it somewhat (in our above-mentioned works, the main elements of the calculation methodology are described and comments to the work \cite{Ryzhov_2005} are given).

This paper presents the results of calculating the angular anisotropy $W(0^{\circ})/W(90^{\circ})$ for the reaction $^{236}$U$(n, f)$ in the neutron energy range from 3 MeV and higher, where the angular anisotropy of fission fragments is described in terms of a statistical model. Note that in this energy region, the dominant contribution to the angular distribution of fragments (\ref{8}) is made only by the term proportional to the coefficient $A_2$, so that the quantity $W(0^{\circ} ) /W(90^{\circ})$ completely describes the angular distribution. The calculation was carried out with the same parameters as the calculation of the fission cross section.

It should be noted that, at high neutron energies, the compound nucleus resulting from the capture of a neutron with a high orbital momentum by the target nucleus has a high spin alignment. This alignment is largely transferred to the nuclei produced by the statistical decay of this ``primary'' compound nucleus. In this case, of course, the longer the chain of decays leading to a given fissioning isotope, the lower its alignment and, consequently, the smaller the angular anisotropy of fission fragments. If, however, a direct or preequilibrium process occurs at the first stage of the reaction, then the resulting residual nucleus has a weak spin alignment. If, therefore, the excitation energy of this nucleus is distributed over all degrees of freedom, i.e. it becomes a ``secondary'' compound nucleus, then the angular distribution of fragments from the fission of such a nucleus is almost isotropic. This, of course, is also true for the fission of the residual nuclei, which are formed at any stage of the statistical decay of such a compound nucleus.

Thus, it can be expected that the smaller the fission cross section through the primary compound nucleus, the weaker will be the angular anisotropy of fission fragments. But this is exactly what is observed: compare the energy dependence of the measured difference $W(0^{\circ})/W(90^{\circ})-1$ (see Fig.~\ref{f15}) and the calculated fission cross section through the primary compound nucleus (dashed curve in Fig.~\ref{f21}) in the region from 30 to 150 MeV, where both quantities tend to zero. Therefore, we calculate only the anisotropy of the fission fragments of the primary compound nucleus and the nuclei formed during its decay. At the same time, we assume that fission fragments of secondary compound nuclei and nuclei formed during their decay are emitted isotropically.

For this calculation, we made only one improvement to our model. In multi-chance fission, the effective moment of inertia ${\cal J}_{\rm eff}$ changes from one fissioning nucleus to another, not only due to the difference in their masses, but also due to different barrier deformations. In addition, ${\cal J}_{\rm eff}$ depends on the energy, since at low excitation energies the moments of inertia of nuclei generally decrease due to the effects of nuclear superfluidity. Previously, we ignored all this and considered ${\cal J}_{\rm eff}$ to be in fact the only fitting parameter that is the same for all nuclei (see \cite{Vorobyev_2020}). However, now for each nucleus the value of $K_0^2$ at the higher barrier is calculated using the equations (\ref{14}) -- (\ref{17}) with the default parameters $\beta_2$.

\begin{figure}
\begin{center}
\includegraphics[scale=0.3]{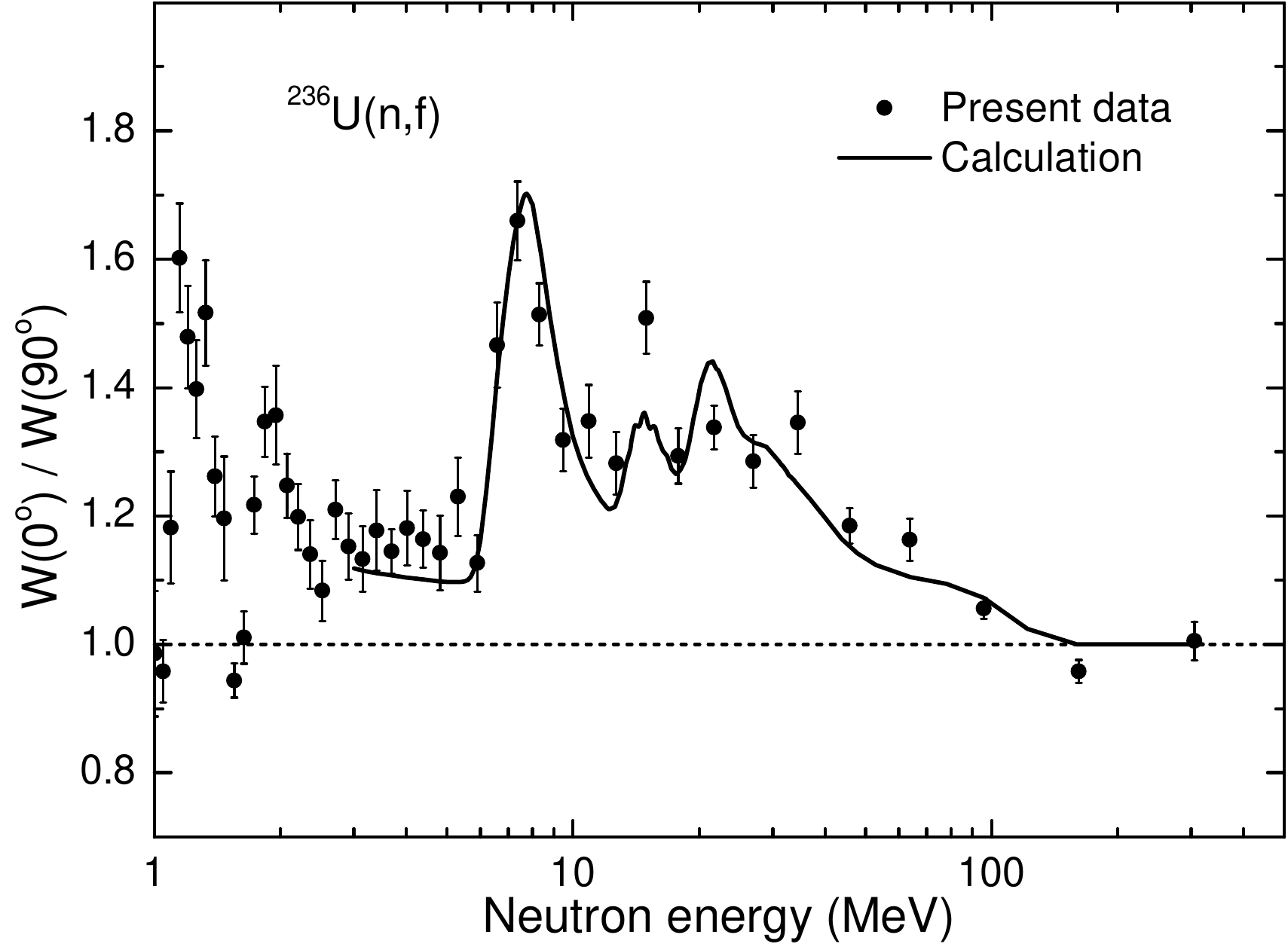}
\caption{The anisotropy of the angular distributions of the $^{236}$U fission fragments compared to the calculation performed for the incident neutron energy above 3 MeV.}
\label{f22}
\end{center}
\end{figure}

The result of the calculation of the angular anisotropy, performed, in fact, without adjusting any parameters, is shown in Fig.~\ref{f22}. On the whole, the calculation reproduces quite accurately both the magnitude of the effect and the characteristic oscillations of the angular anisotropy with increasing neutron energy. The peaks in the angular anisotropy near energies $8$, $14$, $21$ MeV and a slight flattening near $28$ MeV are obviously due to the opening of the 2nd, 3rd, 4th, and 5th chances, respectively (the corresponding peaks in the fission cross section correspond to energies $8$, $16$, $23$ and $32$ MeV).

\section{Conclusion}

The paper presents and describes in detail a technique for simultaneous measurement of the fission cross section ratio and angular distributions of fission fragments of two different nuclei by neutrons depending on the neutron energy, based on the use of position-sensitive multi-wire proportional counters of low pressure. Thus, if the cross-section of one of the nuclei is known, then the described technique allows measuring the cross-section of the other nucleus, as well as the angular distributions of fission fragments of both nuclei.

This technique was used to measure the nuclear fission cross section of $^{236}$U relative to the known nuclear fission cross section of $^{235}$U, as well as the angular distributions of fission fragments of both of these nuclei. The measurements were carried out on the time-of-flight spectrometer of the neutron complex GNEIS at the NRC ``Kurchatov institute'' –- PNPI for neutrons with energies from 0.3 to 500 MeV. The results obtained for the fission cross section of $^{236}$U within the total measurement errors are in good agreement with experimental data obtained earlier by other authors using various techniques and neutron sources. The average statistical accuracy achieved in this work in the energy range above 0.8 MeV is 2.3\%. The average value of the correction, which takes into account the limited solid angle of registration of the position-sensitive fragment detector and the anisotropy of the angular distributions of fission fragments, is $\approx$ 2\%, and its uncertainty is 0.5\%. The total average systematic measurement error is largely determined by the uncertainty of the thickness of the targets used and is 1.4\%.

The results obtained by us on the angular anisotropy of fission fragments of the $^{235}$U and $^{236}$U nuclei by neutrons, depending on the neutron energy, are consistent within the error with the results of previous measurements. The angular anisotropy of fission fragments of $^{236}$U nuclei by neutrons in the energy range above 20 MeV has been obtained for the first time.

The data both on cross sections of nuclear fission by neutrons and on the angular anisotropy of fragments are not only of applied value, but are also of interest for theoretical analysis. The fission cross section measured over a wide range of incident neutron energies, contains fission contributions from a whole chain of isotopes corresponding to a sequence of fission chances, and also, in the region of very high neutron energies, fission contributions from short-lived neutron-deficient isotopes. To demonstrate this, we presented an analysis of the dependence of the fission cross section and the angular distribution of fragments in the $^{236}$U$(n,f)$ reaction on the neutron energy, based on the use of the TALYS-1.9 software package. It is shown that by changing a reasonable number of parameters corresponding to the fission channel, one can obtain a good description of the cross section in the range from 0.5 to $\approx 120$~MeV. However, only the extension of a similar analysis to the fission cross sections of other nuclei can lead to a reliable and consistent set of parameters describing both fission barriers and the level densities of transition states at these barriers.

We also note the success of using the modified TALYS-1.9 complex to describe the energy dependence of the observed angular anisotropy of fission fragments in the $^{236}$U$(n,f)$ reaction in the region of sufficiently high neutron energies. In our opinion, this indicates the validity of our assumption that the decisive role in the formation of the angular anisotropy of the fragments is played by the formation of a primary compound nucleus as a result of the complete fusion of a neutron and a target nucleus. The alignment of the spin of such a compound nucleus due to the contribution of the orbital moment of the captured neutron is quite large. Therefore, both the angular anisotropy of the fission fragments of such a primary compound nucleus and the spin alignment of any of the residual nuclei formed as a result of the cascade decay of this compound nucleus are high. Due to the latter, the angular anisotropy of fragments from fission of such residual nuclei is also high (we are talking about fissions of the 2nd, 3rd, etc. chances). In the modified version of TALYS-1.9, these spin alignments are calculated exactly, which, in combination with the use of the statistical distribution over $K$ (see section~\ref{s04-3}), makes it possible to describe the angular anisotropy of fission fragments practically without involving any adjusting parameters.

This, however, does not at all indicate the uselessness of data on the angular anisotropy of fragments for the theory of fission. Even in the region of validity of the statistical distribution over $K$, the question of the influence of nuclear superfluidity on the effective moments of inertia of fissile nuclei remains open; taking this effect into account can improve the description of the angular anisotropy. Special attention should be paid to the issue of the spin alignment of the residual nuclei arising in direct and preequilibrium processes (we assume that this alignment is small, and therefore we neglect it, but there are no quantitative estimates). And, finally, there is a wide field for studying non-statistical mechanisms of the formation of the angular anisotropy of fragments during the fission of aligned compound nuclei with a low excitation energy, which can provide valuable information on discrete transition states at fission barriers.

\begin{acknowledgments}
The authors express their sincere gratitude to E.M.~Ivanov and the staff of the Accelerator Department of NRC ``Kurchatov Institute'' -- PNPI for the constant friendly support and stable operation of the synchrocyclotron during the experiment as well as L.S Falev for help in creating the experimental setup.

\end{acknowledgments}


\end{document}